\documentclass[lettersize,journal]{IEEEtran}
\AtBeginDocument{%
  }

\usepackage{amsmath}
\usepackage{acro}
\usepackage{tabularx}
\usepackage{booktabs}
\usepackage{csquotes}
\usepackage{wasysym}
\usepackage{multirow}
\usepackage{pdflscape}
\usepackage{multicol}
\usepackage{arydshln}
\usepackage{adjustbox}
\usepackage{subcaption}
\usepackage{array}
\usepackage{xspace}
\usepackage{booktabs} 
\usepackage{longtable} 
\usepackage{bm}
\usepackage{hyperref}
\usepackage{algorithm}
\usepackage{algorithmicx}
\usepackage{algpseudocode}
\usepackage{amsfonts}
\usepackage{xcolor}
\usepackage{float} 
\usepackage{makecell}
\usepackage{caption}
\usepackage{subcaption}
\usepackage{authblk}

\usepackage[inline]{enumitem}
\setlist[itemize]{leftmargin=*, itemindent=8pt}
\setlist[enumerate]{leftmargin=0pt,itemindent=16pt,label=\roman*)}

\usepackage[english]{babel}
\usepackage{listings}
\lstdefinelanguage{json}{
    basicstyle=\footnotesize,
    numbers=none,
    numberstyle=\scriptsize,
    stepnumber=1,
    numbersep=8pt,
    showstringspaces=false,
    breaklines=true,
    frame=lines,
    backgroundcolor=\color{gray!10},
    morestring=[b]{"}, 
    literate=
     *{0}{{{\color{magenta}0}}}{1}
      {1}{{{\color{magenta}1}}}{1}
      {2}{{{\color{magenta}2}}}{1}
      {3}{{{\color{magenta}3}}}{1}
      {4}{{{\color{magenta}4}}}{1}
      {5}{{{\color{magenta}5}}}{1}
      {6}{{{\color{magenta}6}}}{1}
      {7}{{{\color{magenta}7}}}{1}
      {8}{{{\color{magenta}8}}}{1}
      {9}{{{\color{magenta}9}}}{1}
      {:}{{{\color{black}:}}}{1}
      {,}{{{\color{black},}}}{1}
      {"}{{{\color{blue}"}}}{1}
      {PID}{{{\color{red}PID}}}{3}
      {entity_id}{{{\color{red}entity\_id}}}{9}
      {trade_direction}{{{\color{red}trade\_direction}}}{15}
      {source_datetime}{{{\color{red}source\_datetime}}}{15}
      {exchange}{{{\color{red}exchange}}}{8}
      {crytpocurrency}{{{\color{red}crytpocurrency}}}{14}
      {channel_participants}{{{\color{red}channel\_participants}}}{20}
      {entry_prices}{{{\color{red}entry\_prices}}}{12}
      {target_prices}{{{\color{red}target\_prices}}}{13}
      {stop_loss}{{{\color{red}stop\_loss}}}{9}
      {Short}{{{\color{blue}Short}}}{4}
      {Unspecified}{{{\color{blue}Unspecified}}}{7}
      {VET}{{{\color{blue}VET}}}{4}
      {message_text}{{{\color{red}message\_text}}}{12}{SCAPING300}{{{\color{black}SCAPING300}}}{10}
      {VET}{{{\color{blue}VET}}}{3}
      {SHORT}{{{\color{blue}SHORT}}}{5}
      {20x}{{{\color{black}20x}}}{3}{Target1}{{{\color{black}Target1}}}{7}
      {Target2}{{{\color{black}Target2}}}{7}
      {Target3}{{{\color{black}Target3}}}{7}
      {Target4}{{{\color{black}Target4}}}{7}
      {Target5}{{{\color{black}Target5}}}{7}
      {Target6}{{{\color{black}Target6}}}{7}
      {Target7}{{{\color{black}Target7}}}{7}
      {Target8}{{{\color{black}Target8}}}{7}
}
\newfloat{lstfloat}{tbp}{lop}
\floatname{lstfloat}{Listing}

\newcommand{\bi}{\begin{itemize}}
\newcommand{\ei}{\end{itemize}}
\newcommand{\be}{\begin{enumerate}}
\newcommand{\ee}{\end{enumerate}}

\newcommand{\oldstuff}[1]{}
\newcommand{\info}[1]{}
\newcommand{\old}[1]{}
\newcommand{\optional}[1]{}
\newcommand{\consider}[1]{}
\newcommand{\moved}[1]{}
\newcommand{\comments}[1]{}
\newcommand{\temp}[1]{}
\renewcommand{\footnotesize}{\scriptsize}

\makeatletter
\let\disable\@secondoftwo
\addto\extrasenglish{%
  \renewcommand{\sectionautorefname}{\S\@gobble}%
}
\addto\extrasenglish{%
  \renewcommand{\subsectionautorefname}{\S\@gobble}%
}
\addto\extrasenglish{%
  \renewcommand{\subsubsectionautorefname}{\S\@gobble}%
}
\addto\extrasenglish{%
  \renewcommand{\paragraphautorefname}{\S\@gobble}%
}
\makeatother

\DeclareAcronym{osn}{
  short = OSN,
  long  = online social network
}
\DeclareAcronym{gnn}{
  short = GNN,
  long  = graph neural network
}

\DeclareAcronym{gat}{
  short = GAT,
  long  = Graph Attention Networks
}

\DeclareAcronym{gcn}{
  short = GCN,
  long  = Graph Convolutional Networks
}

\DeclareAcronym{sage}{
  short = GraphSAGE,
  long= GraphSAGE
}

\DeclareAcronym{lstm}{
  short = LSTM,
  long  =  Long short-term memory
}

\DeclareAcronym{fca}{
  short = FCA,
  long  =  Financial Conduct Authority
}

\DeclareAcronym{ner}{
  short = NER,
  long  = Named Entity Recognition
}

\DeclareAcronym{cs}{
  short = CS,
  long  = cosine similarity
}

\DeclareAcronym{roc}{
  short = ROC AUC,
  long  = area under the receiver operating characteristic curve
}

\DeclareAcronym{dani}{
  short = DANI,
  long  = Diffusion Aware Network Inference Algorithm
}
\DeclareAcronym{mcc}{
  short = MCC,
  long  = Matthews Correlation Coefficient
}
\DeclareAcronym{sota}{
  short = SOTA,
  long  =  state-of-the-art
}

\begin{document}

\title{\textsc{Perseus}: Tracing the Masterminds Behind Cryptocurrency Pump-and-Dump Schemes}

\author{Honglin Fu, Yebo Feng, Cong Wu, Jiahua Xu}

\maketitle
\begin{abstract}

Masterminds are entities organizing, coordinating, and orchestrating cryptocurrency pump-and-dump schemes, a form of trade-based manipulation undermining market integrity and causing financial losses for unwitting investors.
Previous research detects pump-and-dump activities in the market, predicts the target cryptocurrency, and examines investors and \ac{osn} entities. However, these solutions do not address the root cause of the problem. There is a critical gap in identifying and tracing the masterminds involved in these schemes. In this research, we develop a detection system \textsc{Perseus}, which collects real-time data from the \acs{osn} and cryptocurrency markets. \textsc{Perseus} then constructs temporal attributed graphs that preserve the direction of information diffusion and the structure of the community while leveraging \ac{gnn} to identify the masterminds behind pump-and-dump activities. 
Our design of \textsc{Perseus} leads to higher F1 scores and precision than the \ac{sota} fraud detection method, achieving fast training and inferring speeds. Deployed in the real world from February 16 to October 9 2024, \textsc{Perseus} successfully detects $438$ masterminds who are efficient in the pump-and-dump information diffusion networks.
\textsc{Perseus} provides regulators with an explanation of the risks of masterminds and oversight capabilities to mitigate the pump-and-dump schemes of cryptocurrency.

\vspace{5mm} 

\end{abstract}


\section{Introduction}
\label{sec:intro}
In the cryptocurrency market, a pump-and-dump is a trade-based manipulation tactic in which individuals or groups artificially inflate the price of a cryptocurrency to sell it for a profit, ultimately causing significant losses for investors~\cite{Hamrick2019TheSchemes, Xu2019, trade_based}. In 2023 alone, pump-and-dump manipulators swindled $\$241.6$ million in profit through decentralized exchanges~\cite{chainalysis}, which account for only about $10\%$ of the total trading volume compared to centralized exchanges~\cite{theblock}. This highlights the more substantial impact of pump-and-dump activities on centralized exchanges, drawing the attention of regulators.

\acs{osn}s are largely exploited by pump-and-dump schemes.
The \textit{spreaders}---OSN entities such as Telegram channels, X accounts, chatbots, etc.---disseminate pump-and-dump messages and coordinate investors to buy a specific cryptocurrency collectively through exchanges. As a result of this coordinated buying, the price of cryptocurrency is artificially inflated, allowing some early investors to sell their holdings for profit~\cite{Kamps2018a}. Spreaders can be classified into two categories based on their role in the scheme: \textit{masterminds} and \textit{accomplices}. Masterminds broadcast crowd-pump messages, and accomplices further propagate them. Normally, masterminds campaign for the pump-and-dump and dominate the entire operation, occupying upstream positions in the information diffusion process. In contrast, accomplices follow the lead of the masterminds and retransmit messages to as many investors as possible, forming a community around their respective masterminds~\cite{Ardia2024TwitterPump-and-dumps}.

Previous research has explored the issue of cryptocurrency pump-and-dump schemes, yet a fundamental solution remains elusive. Various rule-based and machine-learning methods have been used to detect and predict pump-and-dump schemes in exchanges~\cite{Kamps2018a, Nilsen2019, LaMorgia2020, Chadalapaka2022, Xu2019, Hu2022}. However, such research does not provide actionable strategies to mitigate the effects of detected or predicted pump-and-dump events. In a notable study, Chen et al.~\cite{Chen2019c} applied an enhanced apriori algorithm to identify exchange users involved in pump-and-dump schemes, suggesting banning involved accounts' transactions. However, this approach risks affecting legitimate transactions and does not prevent masterminds from luring new accounts to continue their fraudulent activities. More recently, researchers have begun linking \ac{osn} entities with pump-and-dump spreaders~\cite{Yahya2024, Merkley2024Crypto-influencers, Ardia2024TwitterPump-and-dumps} through empirical market analyses tied to user activities. This line of research points toward the necessity of identifying the masterminds behind these schemes; without such identification, efforts to mitigate pump-and-dumps by targeting numerous spreaders remain incomplete and less effective.

This study introduces \textsc{Perseus}, a system that fundamentally addresses the issue of pump-and-dump schemes by tracing the masterminds within \acs{osn}s. \textsc{Perseus} is designed to pinpoint masterminds effectively and efficiently, offering a solution to mitigate such schemes with minimal collateral damages and adherence to high ethical standards. To this end, we have developed temporal attributed graph networks to discern masterminds from accomplices. The deployment of \textsc{Perseus} involves several key steps. Initially, we scrape real-time data from Telegram channels and the cryptocurrency market. Subsequently, we construct temporal graphs while incorporating \ac{osn}, topological, and market features. In the final step, we apply \ac{gnn} to temporal attributed graphs to detect masterminds.

To deploy \textsc{Perseus}, we use real-world data scraped from April 2018 to February 2024 for evaluation. During this period, $660$ cryptocurrencies and $2,103$ Telegram channels have been monitored by \textsc{Perseus}. We employ regular expressions for the $27,365,232$ messages, extracting the $733,128$ pump-and-dump-related messages. By grouping messages, we compile $4,101$ instances of pump-and-dump events. \textsc{Perseus} then processes these instances, generating $920$ temporal attributed graphs with $9,666$ nodes and $23,293$ edges. We evaluate \textsc{Perseus}'s ability to identify masterminds and find \textsc{Perseus} beats the benchmark in F1 score, precision, accuracy, and \ac{mcc} and demonstrates rapid training and stable inference speeds. In addition, we document the characteristics of masterminds detected and present two case studies to illustrate how \textsc{Perseus} detects masterminds. From February 16 to October 9 2024, \textsc{Perseus} successfully identified $290$ masterminds involved in pump-and-dump activities, potentially reducing the related trading volume impacts by approximately $\$3.24$ trillion.

Our paper makes the following contributions:
\begin{itemize}[leftmargin=*]
    \item We are the first to investigate the masterminds of cryptocurrency pump-and-dump and offers a framework \textsc{Perseus} detecting $438$ masterminds across $322$ cryptocurrencies, with high efficacy and efficiency.
    
    \item \textsc{Perseus} assembles data from the \acs{osn} and cryptocurrency market and processes an unprecedented dataset of cryptocurrency pump-and-dump schemes from $2,103$ channels, exceeding the scope of prior studies that examined between 50 to 700 channels \cite{Hu2022, Mirtaheri2021, Nghiem2021, Xu2019}.

    \item We allow for an explanation of the risk of masterminds. Masterminds are efficient in their information diffusion networks, and accomplices get crowd-pump information directly from them.
    
\end{itemize}

\section{Background}
\label{sec:backgroud}

\subsection{Pump-and-dump Schemes}
In the cryptocurrency market, a pump-and-dump is a form of market manipulation similar to a trade-based pump-and-dump in traditional financial markets, where individuals or groups purchase a cryptocurrency, artificially inflating its price through strategic trading rather than spreading false rumors, and subsequently sell it for profit~\cite{trade_based}. Importantly, all illicit pump activities include a dumping phase, during which the sell-off begins, often leading to a price drop and financial losses for unsuspecting investors.

\begin{figure}[!t] 
    \centering
    \begin{minipage}{0.49\linewidth}
        \centering
        \includegraphics[width=\linewidth]{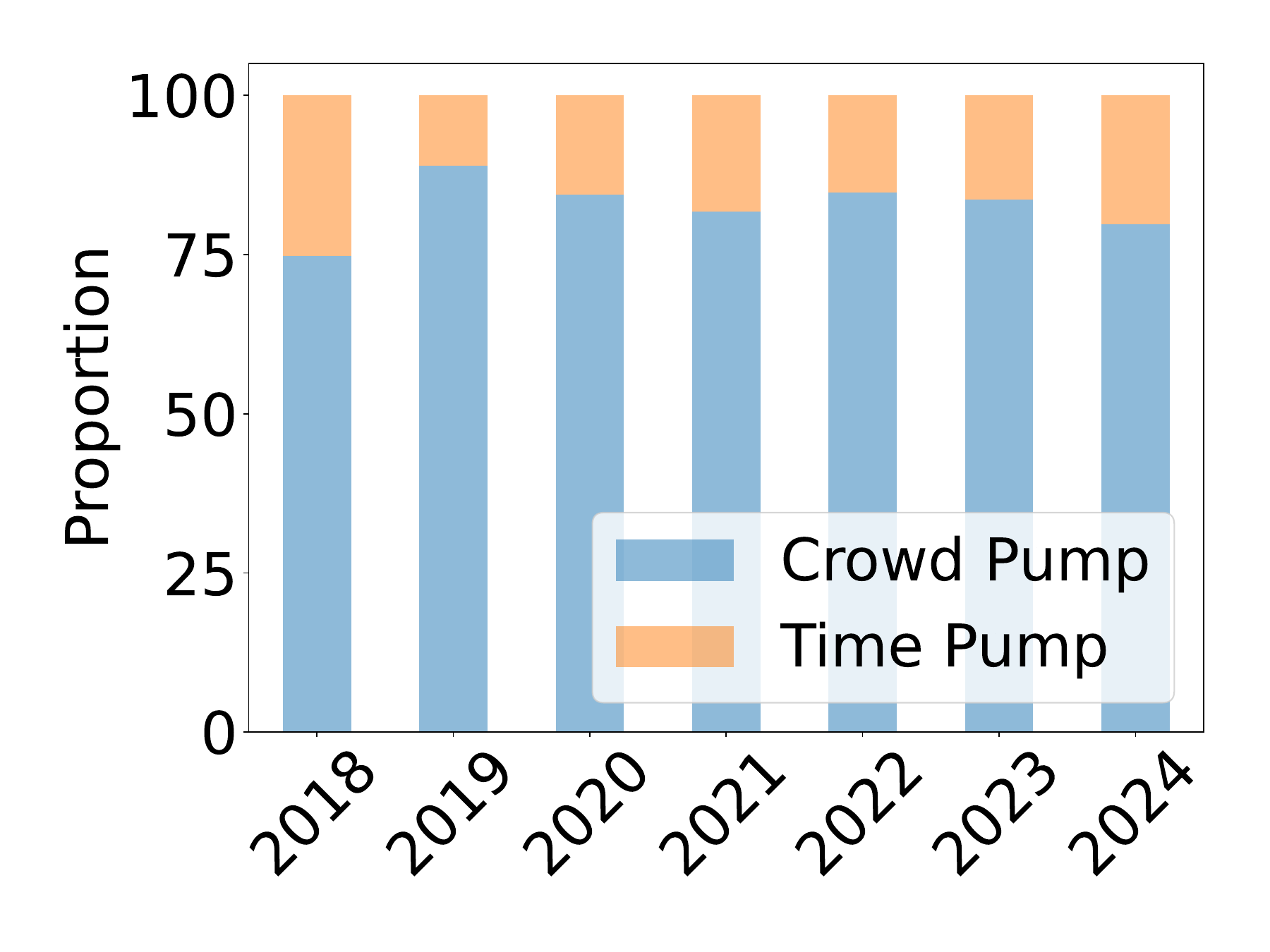}
        \subcaption{Proportion of Pumps Over Time.}
        \label{fig:Clustering_Coefficient}
    \end{minipage}%
    \hfill
    \begin{minipage}{0.49\linewidth}
        \centering
        \includegraphics[width=\linewidth]{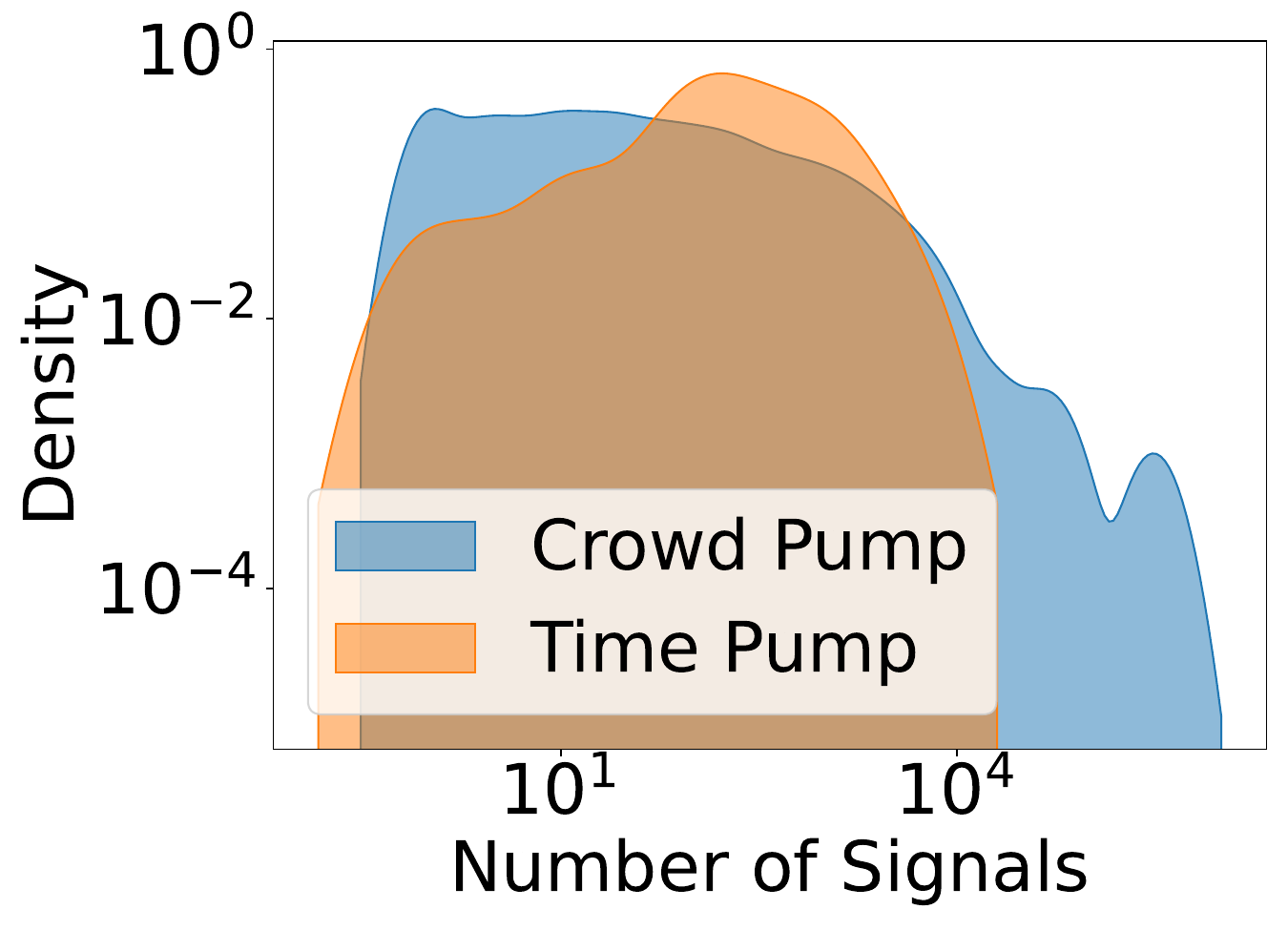}
        \subcaption{Kernel Density of Pumps per Channel.}
        \label{fig:Efficiency}
    \end{minipage}
    \caption{\textsc{Perseus} assembles the most comprehensive pump-and-dump data with $2103$ pump-and-dump channels, showing that crowd-pumps are more prevalent than time-pumps.}
    \label{fig:comparison}
\end{figure}

\begin{figure*}[!t] 
\centering
\begin{minipage}{0.33\linewidth} 
  \centering
  \includegraphics[height=4cm]{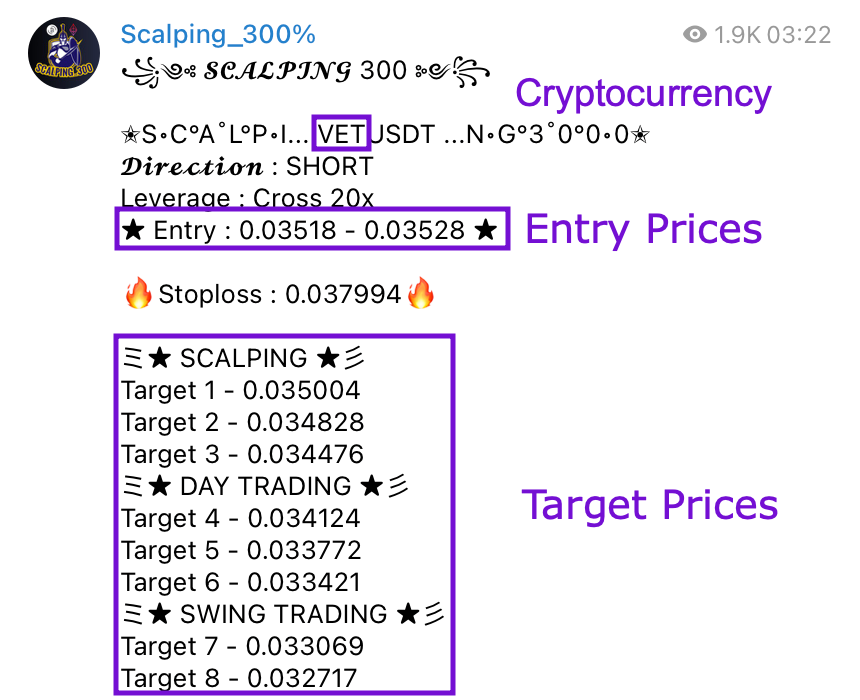} 
  \subcaption{The crowd-pump message coordinates investors to purchase the cryptocurrency when the price is within the entry prices and sell them at the target prices.}
  \label{subfig:crowd_pump message}
\end{minipage}\hfill 
\begin{minipage}{0.65\linewidth} 
  \centering
  \includegraphics[height=4cm]{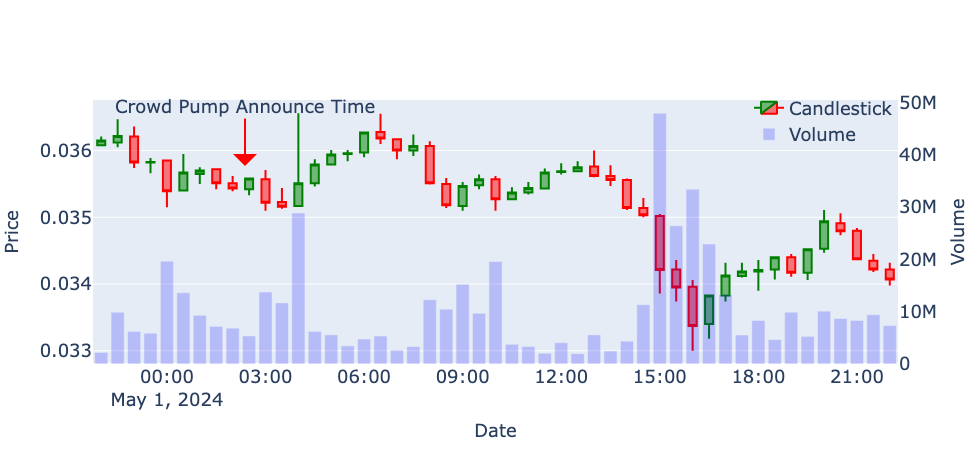} 
  \subcaption{The market reaction to the crowd-pump message in \autoref{subfig:crowd_pump message}.}
  \label{subfig:leverage_defi}
\end{minipage}
\caption{A crowd-pump on \textit{VET}, the native token for Vechain, a public blockchain secured by a proof of authority consensus mechanism. Note that the example shows a pump-and-dump on a short contract, a financial instrument profiting from a decline in the value of an asset. In this case, masterminds encourage investors to buy the short contract and sell the spot cryptocurrency.}
\label{fig:crowdmarket}
\end{figure*}

Pump-and-dump schemes, which manipulate targeted cryptocurrencies through coordinated buying and selling, are divided into time-pumps, which synchronize trading activities~\cite{Xu2019, Hu2022}, and crowd-pumps, which organize trades through predefined price boundaries~\cite{Morgia2022, Hamrick2021}. Because crowd-pumps do not need precise timing coordination, they are more prevalent than time-pumps, as depicted in \autoref{fig:comparison}. The crowd-pump is a pump whose messages have prices specifying when to dump the cryptocurrency. \autoref{fig:crowdmarket} provides an example of a crowd-pump, including its messages, characterized by the specified trading pair and the predefined buying and selling prices, and its market impact.

\subsection{Crowd-Pump and \acs{osn}}

\ac{osn}s are increasingly exploited by spreaders to enhance the effects of crowd-pumps. Specifically, on platforms like Telegram, spreaders establish channels where only they can talk to coordinate investors to align their trades with the crowd-pump messages. Masterminds broadcast crowd-pump messages and accomplices further propagate them, thus forming a community around their respective masterminds~\cite{Ardia2024TwitterPump-and-dumps}. This process mirrors the diffusion of information, where crowd-pump messages diffuse from masterminds to their accomplices, resulting in a coordinated crowd-pump event.


\subsection{Regulatory Action}
\label{regulatory}

The prevalence of crowd-pumps messages across \ac{osn} have drawn attention from regulators. Authorities such as the Financial Conduct Authority, the European Union, and the U.S. Senate are exploring methods to counteract the impact of pump-and-dump schemes on \ac{osn}. The Financial Conduct Authority defines financial crimes as criminal activities related to money, financial services, or markets, including fraud and market misconduct \cite{FCA}. The European Union assesses fraudulent cryptocurrency marketing and defines market manipulation as deceptive trading, biased media statements, and false information to influence prices \cite{euregulation}. Meanwhile, the U.S. is considering legislation to clarify regulatory jurisdiction over cryptocurrency \cite{bill}.

\section{System Design}
\label{sec:system_design}

\subsection{System Overview}
\begin{figure*}[!t]
  \centering
  \includegraphics[width=\textwidth]{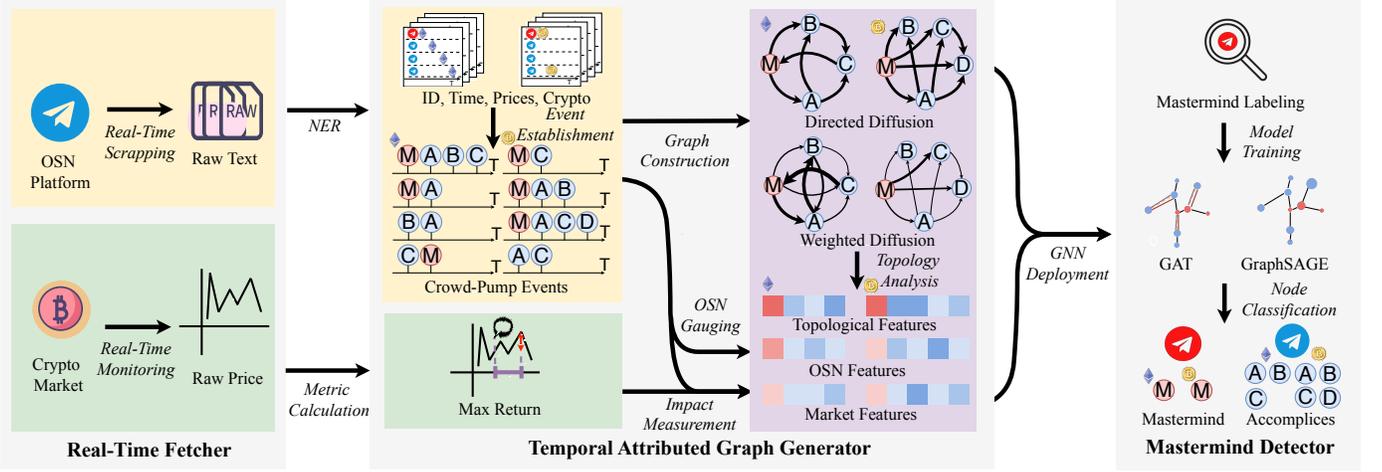}
  \caption{\textsc{Perseus} consists of three components: real-time fetcher, temporal attributed graph generator, and mastermind detector. The real-time fetcher collects data from \ac{osn} and the cryptocurrency market, which is processed by temporal attributed graph generators to create information diffusion graphs that serve as input for the mastermind detector to identify masterminds. The real-time fetcher and temporal attributed graph generator concurrently process data from \ac{osn} and cryptocurrency markets.}
  \label{fig:pipeline}
\end{figure*}

Since crowd-pumps are much more prevalent than time-pumps, we design \textsc{Perseus} based on the characteristics of crowd-pumps to detect masterminds. We design \textsc{Perseus}' pipeline as three integral components: real-time fetcher, temporal attributed graph generator, and mastermind detector. The real-time fetcher gathers data, which is passed to the temporal attributed graph generator for producing the information diffusion graphs and node features. Finally, the mastermind detector builds temporal attributed graph networks to classify spreaders into masterminds and accomplices. In particular, the real-time fetcher and temporal attributed graph generator parallel mine data from \ac{osn} platforms and cryptocurrency markets. An overview of \textsc{Perseus} is presented in \autoref{fig:pipeline}.  

\subsection{Real-time Fetcher}
We collaborate with Cloudburst \cite{cd} to deploy the real-time fetcher. The real-time fetcher is divided into two parallel streams: one focusing on the cryptocurrency market and the other on \acs{osn} platforms. 

\subsubsection{\acs{osn} Real-time Scrapping}

Previous research shows that Telegram~\cite{Telegram} hosts the most pump-and-dump groups and messages compared to other \acs{osn}s~\cite{Xu2019}. Consequently, we choose Telegram as the primary source to collect \acs{osn} data. To collect data from Telegram, we search for keywords commonly found in the names of pump-and-dump groups in a disclosed database published by Morgia~\cite{LaMorgia2020}. The terms \enquote{pump} and \enquote{signal} emerge as the most prevalent. As a result, we employ search queries \enquote{crypto pump Telegram} and \enquote{crypto signal Telegram} on search engines Google and Bing.

The initial search yields 216 channels. We then deploy bots to scrape real-time raw text messages from each channel, including invitation links to additional Telegram channels. This facilitates an iterative process in which following these links expands our channel list. Through this method, we assemble a collection of $2,103$ channels. The bots also collect metadata for each channel, including the timestamp of each message and the channel identifier. 

We acknowledge that our search strategy might overlook some Telegram channels and other social media platforms. However, we are confident that this approach helps us identify the most representative and popular channels. With their highest numbers of group members and messages, the Telegram channels we've collected are sufficient to find the masterminds.

\subsubsection{Cryptocurrency Market Real-time Monitoring} 

For each cryptocurrency, we retrieve trade-by-trade price data covering three days before and after each crowd-pump message. We choose three days as the window size because the reported duration of the crowd-pump messages usually will not exceed three days. Given the substantial volume of trades with Bitcoin and Ethereum, we capture data at one-minute intervals to reflect market reactions to crowd-pumps.


\subsection{Temporal Attributed Graph Generator}
The temporal attributed graph generator processes both \ac{osn} and cryptocurrency data like the real-time fetcher. For the \ac{osn} data, we extract crucial information from crowd-pump messages through \ac{ner}. Then, we establish crowd-pump events based on time gaps used to construct information diffusion graphs. In parallel, for the cryptocurrency market data, we calculate the maximum return for each crowd-pump. Ultimately, we engineer market, \ac{osn}, and topological features with the processed data.

\subsubsection{\acs{osn} Processing}

\lstset{
    language=json,
    stringstyle=\color{blue},
    numberstyle=\color{magenta},
    frame=single,
    rulecolor=\color{black},
    captionpos=b,
    breaklines=true,
    breakatwhitespace=true,
    postbreak={},  
    columns=fullflexible,
    numbers=none  
}

\captionsetup[lstfloat]{font={small,bf}}

\begin{lstfloat}[!b] 
\begin{lstlisting}[language=json,caption={\small NER extracted JSON data for the crowd-pump in \protect\autoref{fig:crowdmarket}. PID is the crowd-pump message ID. Entity ID identifies the broadcasting channel. Trade direction indicates position. Source datetime is the message timestamp. Exchange and commodity specify where and what is pumped. Channel participants show group size. Entry, target, and stop-loss prices are instructional trade levels.},label=json_sample]
[
  {
    "PID": 398868,
    "entity_id": "-1001313911314",
    "trade_direction": "Short",
    "source_datetime": "05-01-2024 03:14:42",
    "exchange": "Unspecified",
    "crytpocurrency": "VET",
    "channel_participants": 38046,
    "entry_prices": [0.03518, 0.03528],
    "target_prices": [0.035004, 0.034828, 0.034476, 0.034124, 0.033772, 0.033421, 0.033069, 0.032717],
    "stop_loss": 0.037994,
    "message_text": "SCAPING300. VETUSDT. Direction: SHORT. Leverage: Cross 20x. Entry: 0.03518, 0.03528. Stoploss: 0.037994. SCALPING: Target1 - 0.035004, Target2 - 0.034828, Target3 - 0.034476. DAY TRADING: Target4 - 0.034124, Target5 - 0.033772, Target6 - 0.033421. SWING TRADING: Target7 - 0.033069, Target8 - 0.032717"
  }
]
\end{lstlisting}
\end{lstfloat}

\paragraph{Named Entity Recognition}
The raw text data from the \acs{osn} channels undergo a \ac{ner} process using regular expressions. \ac{ner} is an information extraction technique to identify predefined semantic types~\cite{Li2022ARecognition}. We focus on extracting key details such as timestamps, channel identifiers, names of cryptocurrencies, entry prices, target prices, and the recommended trading positions (either long or short). \autoref{json_sample} shows an example of this structured data in JSON format.

\paragraph{Crowd-pump Event Establishment}
A crowd-pump event is a series of crowd-pumps. We establish crowd-pump events based on the time gaps between messages. To achieve this, we group messages observed within a period of time according to cryptocurrency and trade direction. Within each group, messages are segmented into events using a time gap threshold, the lesser of the 95th percentile of time gaps of the group, or a three-day cap, as reported crowd-pumps rarely last longer. If a spreader sends multiple messages in an event, only the first is retained. Messages in the same event may have different entry or target prices.

\paragraph{Crowd Pump Events}
A crowd-pump labeled $c$ includes a detailed message. Each message is associated with the target cryptocurrency $x$, the time $t$ at which it was posted, and the identity $v$ of the spreader who sent it. Thus, we denote a crowd-pump as a tuple $(v^x,t^x)$. When a crowd-pump message is sent on an \acs{osn}, other spreaders read and retransmit it, passing on the information. We refer to this process of crowd-pump information diffusion as a crowd-pump event $d$. Let $T$ be a continuous time interval in which crowd-pump messages are observed where $T = [t_1, t_n]$. A crowd-pump event $d$ targeting cryptocurrency $x$ can be formalized as $d^x_T =  \{(c_1), (c_2), \ldots, (c_n)\} = \{(v^x_1, t^x_1), (v^x_2, t^x_2), \ldots, (v^x_n, t^x_n) \}$. Let the overall observation period $\tau$ be the union of several disjoint or contiguous time intervals where $\tau = T_1 \cup T_2 \cup \ldots \cup T_n$. Each $T_i$ is a continuous time interval during which crowd-pump messages are collected. For the target cryptocurrency $x$, the collection of crowd-pump events over the entire period $\tau$ is defined as: $D^x_\tau = \{d^x_{T_1},d^x_{T_2}, \ldots, d^x_{T_n} \}$, where each $d^x_{T_i}$ is the crowd-pump events observed in the time interval $T_i$.

\begin{figure}[!b]
  \centering  \includegraphics[width=\linewidth]{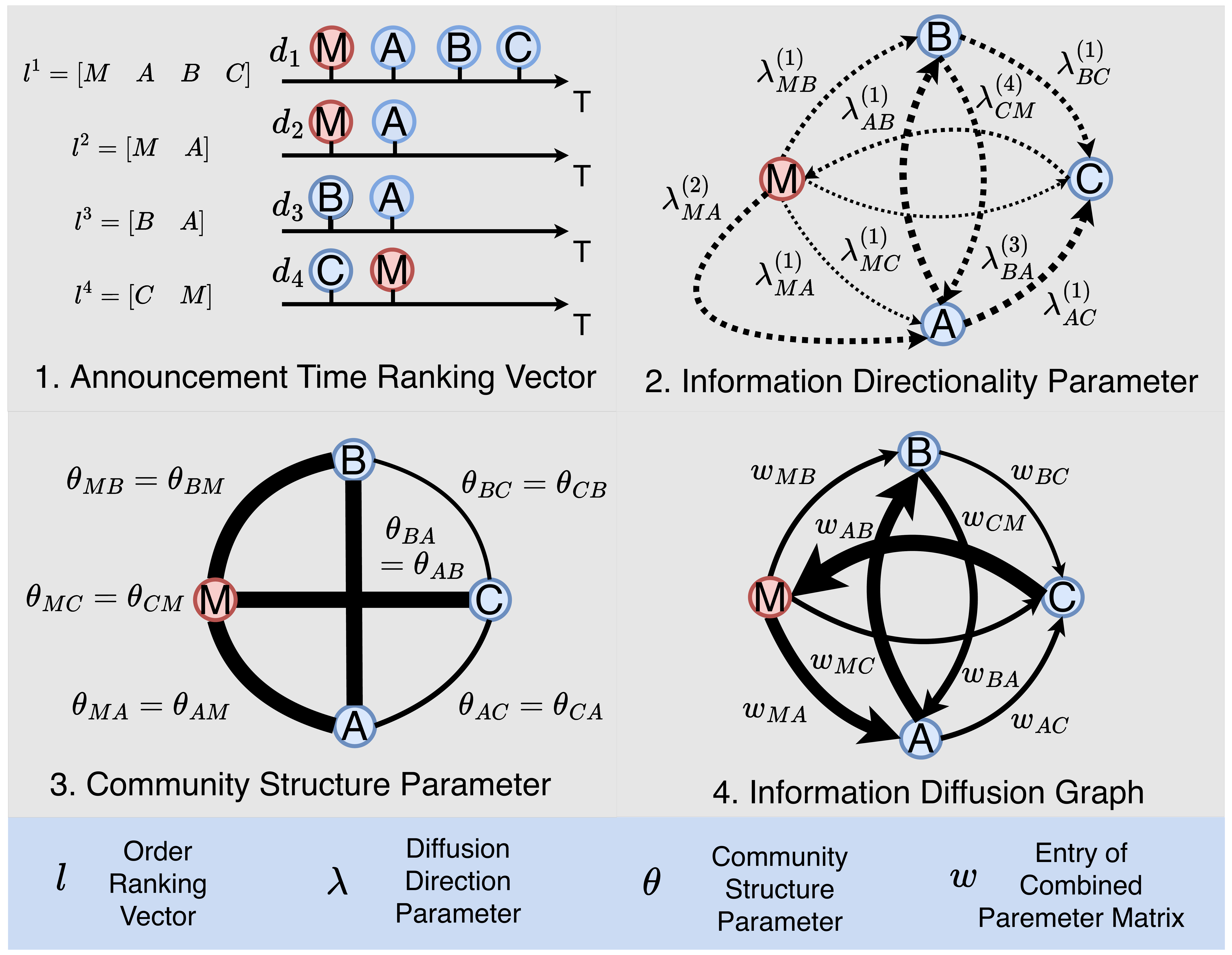}
  \caption{Illustration of graph construction for Ethereum spreaders in \autoref{fig:pipeline}. In step 1, we sort the crowd-pump messages chronologically. In step 2, we derive the direction of information diffusion. In step 3, we infer the community structure. In step 4, we integrate the parameters of community structure and information direction together.}
  \label{fig:graph_infer}
\end{figure}

\paragraph{Graph Construction}
Let $V$ represent the set of spreaders divided into masterminds $V_m$ and accomplices $V_n$, such that $V = V_m \cup V_n$. Define $V_m$ as the set of masterminds $\{v_{m_1}, v_{m_2}, \dots, v_{m_i}\}$ and $V_n$ as the set of accomplices $\{v_{n_1}, v_{n_2}, \dots, v_{n_j}\}$. The spreaders utilize \acs{osn}s to propagate crowd-pumps messages, forming information diffusion pathways as edges $E$ on a graph $G$. Thus, the graph $G = (V, E)$ defines the crowd-pump information diffusion network among spreaders. We construct the crowd-pump spreader network $G$ by inferring the edges $E$ from the patterns of information diffusion $D^x_\tau$.

We construct temporal graphs in each period by customizing the \ac{dani}~\cite{dani}. Temporal graphs, where nodes are entities and temporal edges are relationships between entities over time, are often used to model networks that evolve \cite{Huang2023TemporalGraphs}. \ac{dani} is used to infer the latent network structure by evaluating information diffusion patterns, which are sequences of message propagation~\cite{Ramezani2023JointFactorization, Gao2023Pairwise-interactions-basedCascades}. We choose this algorithm because it preserves the direction of information diffusion and community structure, which is crucial in cryptocurrency trading, where communities shape the information flow.

Over a period $\tau$, we observe a set of crowd-pump events $D^x_\tau$. We infer the structure of the hidden diffusion network $G^*_{D^x_\tau}$ for a cryptocurrency $x$ by maximizing the likelihood of $P(D^x_\tau|G)$ for crowd-pump events occurring over the period $\tau$. After processing all crowd-pump events in $D^x_\tau$ targeting cryptocurrency $x$ over the period $\tau$, we get dynamic graphs $G_{D^{x}_{\tau}}$.

$G_{D^x_\tau}=(E(D^x_\tau), V(D^x_\tau))$ represents the underlying information diffusion network for crowd-pump events $D^x_\tau$. $V(D^x_\tau)$ is the set of nodes representing the spreaders participating in crowd-pump $D^x_\tau$, and $E(D^x_\tau)$ is the edge set denoting the direction of information diffusion and the structure of the community. We infer the network $G_{D^x_\tau}$ by maximizing the likelihood of message passing in crowd-pump events $D^x_\tau$ through the Markov chain. Adapting the proof of \ac{dani}, we prove $G' \propto \underset{G}{\mathrm{argmax}} \left( \sum_{(r,s) \in G} \sum_{d_i \in D^x_\tau} \left(\frac{\theta_{rs}}{\lambda^{(i)}_{rs}} \right) \right)$.
Then we compute the adjacent matrix by maximizing the likelihood of the above:
$G' = \underset{G | 0 \leq W \leq 1}{\mathrm{argmax}}\ P(D^x_\tau|G)$, where $w_{rs}=\frac{\theta_{rs}}{\lambda_{rs}}$ is a weighted adjacent matrix. $\theta_{rs} = \frac{|\mathcal{I}_n(r) \cap \mathcal{I}_n(s)|}{|\mathcal{I}_n(r) \cup \mathcal{I}_n(s)|}$ and 
$\lambda^{(i)}_{rs} = \frac{h^{(i)}_{rs}}{\sum_{y \in V_{D^{x}_{\tau}
}} h^{(i)}_{ry}}$ are two parameters used to preserve the structure of the community and the directionality of the diffusion of information. $\mathcal{I}_n$ in $\theta$ is used to calculate the Jaccard similarity coefficient between two spreaders $r$ and $s$. For each crowd-pump, $h$ in $\lambda$ is computed using the spreader ranking vector sorted on the time of the announcement of the message, $l$. Specifically, $ h^{(i)}_{rs} = [l^{(i)}_{s} \times (l^{(i)}_{s} - l^{(i)}_{r})]^{-1}$.

\autoref{fig:graph_infer} outlines the computation of the adjacency matrix. In Step 1, ranking vectors \( l \) are created by sorting message times of spreaders during crowd-pump events, capturing the diffusion order. Step 2 quantifies information directionality using parameters \( h \) and \( \lambda \), where \( h \) represents the diffusion strength between spreaders, and \( \lambda \) adjusts \( h \) based on the number of connections a spreader has, ensuring influence is proportionally distributed. Step 3 determines community structure through parameters \( \mathcal{I}_n \) and \( \theta \), measuring event participation and spreader overlap. Step 4 integrates \( \lambda \) and \( \theta \) to compute diffusion weights \( w \), combining community structure and directionality.

We construct two temporal attributed graphs for each cryptocurrency $x$ during the period $\tau$: a weighted diffusion graph and a directed diffusion graph. The weighted diffusion graph is the adjacent matrix $W_{rs}$, and directed diffusion $W^*_{rs}$ is created as follows: $W^*_{rs} = 
    \begin{cases} 
        1 & \text{if } W_{rs} > W_{sr} \\
        0 & \text{otherwise.}
    \end{cases}$. The matrix $W^*_{rs}$ is constructed by comparing the probabilities of diffusion from one spreader to another and designating the direction of the edge from the spreader with a higher probability of sending messages to the lesser one, capturing the main direction of diffusion of information between spreaders.

\subsubsection{Market processing}

\paragraph{Max Return}
To assess the impact of each crowd-pump message, we calculate the maximum return within three days after the announcement. This is the return from the price at the announcement time to the highest price within this window, assuming an immediate market reaction. The three-day window is chosen because crowd-pump effects typically do not last longer.

\subsubsection{Node Features}
With the \ac{ner} extracted data from \acs{osn}, the return of each message, and constructed graphs, we engineer the \acs{osn}, market, and topological features for each spreader within $\tau$.

\paragraph{\acs{osn} Feature}
We define two features: \textbf{total targets achieved}, the number of target prices met per spreader, and \textbf{rating}, the ratio of targets achieved to the total number of targets within $\tau$. A target is considered achieved if its price falls below the maximum price within three days of the announcement.

\paragraph{Market Features}
We engineer the market feature \textbf{average increase}, the average return of all crowd-pump messages a spreader sends within $\tau$.
\begin{table*}[t] 

\centering 
\caption{Features for unweighted and weighted graphs for node (\(v\)) in its ego network. Variables include number of nodes (\(n\)) and others as specified in the descriptions.} 
\resizebox{\textwidth}{!}{%

\footnotesize
\begin{tabular}{lcccc} 
\toprule 
\textbf{Ego Network} & \textbf{Feature} & \textbf{Unweighted Formula} & \textbf{Weighted Formula} & \textbf{Description of Variables} \\ 
\midrule 
\multirow{5}{*}{Degree-out} & Effective size & \(n - \frac{1}{n} \sum t_i\) & \(n - \frac{1}{n} \sum w_i\) & \multirow{2}{*}{\makecell{\(t_i\): the number of ties that alter $i$ has with other alters within the ego network of node $v$ \\ \(w_i\): the sum of weights of all outwards edges that a specific alter $i$ has with other alters within the ego network of node $v$}} \\ 
& Efficiency & \(1 - \frac{1}{n^2} \sum t_i\) & \(1 - \frac{1}{n^2} \sum w_i\) & \\ 
\cmidrule(lr){2-5}
& Out degree & \(q\) & \(Q\) & \multirow{2}{*}{\makecell{$q$: total one-degree out edges for ego $v$ \\ $Q$: total sum of one-degree out weights of edges for ego $v$}} \\ 
& Out-ratio & \(\frac{q}{n}\) & \(\frac{Q}{n}\) & \\ 
\cmidrule(lr){2-5}

& Density & \( \frac{2m}{n(n-1)}\) & \( \frac{2M}{n(n-1)}\) & \makecell{\(m\): total unweighted edges for the ego network of $v$ \\ \(M\): total weight sum for the ego network of $v$}\\ 
\midrule 
Degree-in & In-ratio & \(\frac{\text{indeg}(v)}{n}\) & \(\frac{\text{indeg}_w(v)}{n}\) & 
\makecell{
\(\text{indeg}(v)\): unweighted in-degree\\ \(\text{indeg}_w(v)\): weighted in-degree}\\ 
\bottomrule 
\end{tabular} 
}

\label{tab:centrality_measures} 
\end{table*}

\paragraph{Topological Features}
Over period $\tau$, we construct directed and weighted information diffusion graphs for cryptocurrency $x$ and engineer unweighted and weighted features for each spreader. \textbf{Clustering coefficient}, \textbf{closeness centrality}, \textbf{betweenness centrality}, and \textbf{pagerank} are computed for each node on the whole graph. \textbf{In-ratio}, \textbf{out degree}, \textbf{out-ratio}, \textbf{efficiency}, \textbf{effective size}, and \textbf{density} are calculated in the ego network of each node. An ego network is a network that focuses on a single node (the ego) and the immediate direct connections that node has with other nodes (the alters)~\cite{ego}. For the one-step degree-in ego network, we calculate the in-ratio. For the one-step degree-out ego network, we calculate out degree, out-ratio, efficiency, effective size, and density features. The formulas for these features of the ego network are detailed in \autoref{tab:centrality_measures}.

\subsection{Mastermind Detector}

\subsubsection{Mastermind Labeling}
\begin{figure}[!t]
  \centering
  \includegraphics[width=\linewidth]{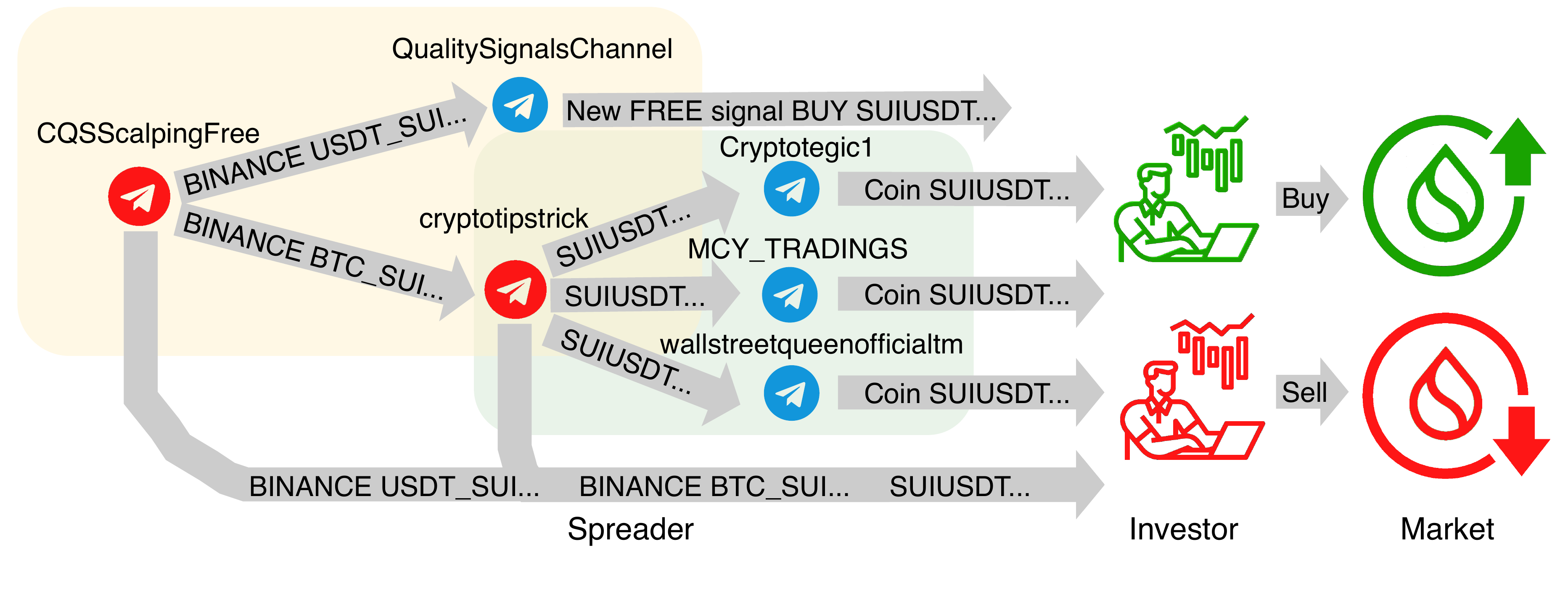}
  \caption{\textit{SUI} crowd-pump illustration. Red Telegram icons represent masterminds, and green ones represent accomplices. Color in block indicate community.  Masterminds broadcast crowd-pump messages to accomplices who further propagate the information, instructing investors to buy and sell cryptocurrency.}
  \label{fig:case_sui_crop}
\end{figure}

To identify the masterminds, we perform financial forensic analyses. We use the Louvain algorithm to analyze community structures, check the main direction of information diffusion, examine each message's market impact, and verify pump timing to see if a suspected mastermind is upstream in crowd-pump events. After identifying potential masterminds, we rule out spreaders if they exhibit inconsistent and fragmented thoughts, unclear purpose, vague and ambiguous statements, inaccuracies, sporadic use of technical terms, humorous and playful tone, random and unrelated references, and lack of grammar and punctuation.

To elaborate the investigation, we construct the underlying crowd-pump information diffusion network in \autoref{fig:case_sui_crop} with the details of the crowd-pump presented in \autoref{tab:SUI}. Our investigation focuses on identifying masterminds behind the cryptocurrency \textit{SUI} during the period from February 6 to February 13 2024. We analyze two crowd-pump events, denoted as $d_1$ and $d_2$. The crowd-pump events involving \textit{SUI} are led by two masterminds, supported by four accomplices.

In \autoref{fig:case_sui_crop}, there are two communities. Focusing on the yellow community, the spreader CQSScalpingFree emerges as a key figure, initiating and participating in $d_1$ and $d_2$ crowd-pump events. Given the extent of its influence in two crowd-pump events, we identify CQSScalpingFree as the mastermind within the yellow community. In the green community, cryptotipstrick follows the lead of CQSScalpingFree in crowd-pump event $d_2$. Although CQSScalpingFree is the initiator, cryptotipstrick exerts a greater influence by influencing three spreaders. This level of influence establishes cryptotipstrick as the mastermind within the yellow community.

For one crowd-pump event, there could be multiple masterminds or none. The first spreader of a crowd-pump event is not necessarily identified as the mastermind, as timing alone is insufficient. The mastermind must also coordinate the accomplices and direct the campaign.

The above selection rules are made because the masterminds are professionals with years of crowd-pump broadcasting records and desire their crowd-pump messages to be clear and effective. Such filtration rules help us design mastermind detection as a node classification task within temporal attributed graphs, as inspired by various studies applying graphs to perform forensics on blockchain~\cite{Song2025IllicitBlockchains, Du2024BreakingLearning, Guo2024Graph-BasedBlockchain, Liu2024FishingData}.

\subsubsection{GNN in Mastermind Detection}
For a cryptocurrency $x$ over the period of time $\tau$, we construct the temporal attributed graph $G_{D^x_\tau}=(E(D^x_\tau), V(D^x_\tau))$. We assume that at least one node, $v_m \in V(D^x_\tau)$, is the mastermind organizing crowd-pump events. Each node $v_i \in V(D^x_\tau)$ is associated with a feature vector $\mathbf{z}_i \in \mathbb{R}^n$. Given the labels $\{y_i\}_{i=1}^{|V|}$, where $y_i$ represents the label of node $v_i$, 1 for mastermind and 0 for accomplices, \textsc{Perseus}' objective is to learn a mapping function $f: \mathbb{R}^d \rightarrow \mathbb{R}$ that assigns a probability score to each node, representing how likely it is to be a mastermind. To achieve that, we apply \acl{gat} (\acs{gat}) \cite{GAT} and \acs{sage}~\cite{graphsage}. These architectures are chosen because we need to classify spreaders for unseen cryptocurrency networks, which suggests that an inductive task model is preferred~\cite{Xu2020InductiveGraphs}. 

\subsubsection{Graph Attention Networks}
\acs{gat} introduces an attention mechanism useful to detect masterminds, enabling the model to focus on key information diffusion pathways. The adaptability of \acs{gat} allows us to emphasize the most relevant spreaders--central to information diffusion--while downplaying the less relevant ones.

In the mastermind detection task, \acs{gat} updates each node’s features by aggregating information from its neighbors, weighed by learned attention coefficients. This enables the network to identify spreaders with significant influence and typical information diffusion patterns of masterminds. The node feature update formula in a \acs{gat} layer is given by:
$h'_i = \sigma\left(\sum_{j \in \mathcal{N}(i)} \alpha_{ij} W h_j\right)$,
where $\alpha_{ij}$ represents the attention weight between node $i$ and its neighbor $j$, and $\mathcal{N}(i)$ is the set of neighbors of node $i$. These learned weights prioritize the most critical nodes for the task, allowing \acs{gat} to differentiate between masterminds and accomplices based on their roles in the graph.

\subsubsection{GraphSAGE}
\acs{sage} aggregates information from a node’s local neighborhood iteratively, generating node embedding that captures both local and global structure. This iterative process is key to detecting masterminds, as it allows nodes to incorporate information from increasingly larger neighborhoods, capturing the extended influence of a mastermind over the network.

In mastermind detection, \acs{sage} can effectively combine features from masterminds' immediate neighbors as well as those farther away, capturing the diffusion influence that masterminds may exert. The update formula for each node’s features in a \acs{sage} layer is: $h'_i = \sigma\left(W \cdot \text{AGGREGATE}\left(\{h_i\} \cup \{h_j, \forall j \in \mathcal{N}(i)\}\right)\right)$,
where $\text{AGGREGATE}$ is an aggregation function. In our case, mean pooling combines the node’s features with those of its neighbors.

\section{Evaluation}
\label{sec:evaluation}
\begin{figure*}[!th]
\centering
\begin{minipage}{\textwidth}
  \centering
  \begin{subfigure}[b]{0.19\linewidth}
    \centering
    \includegraphics[width=\linewidth]{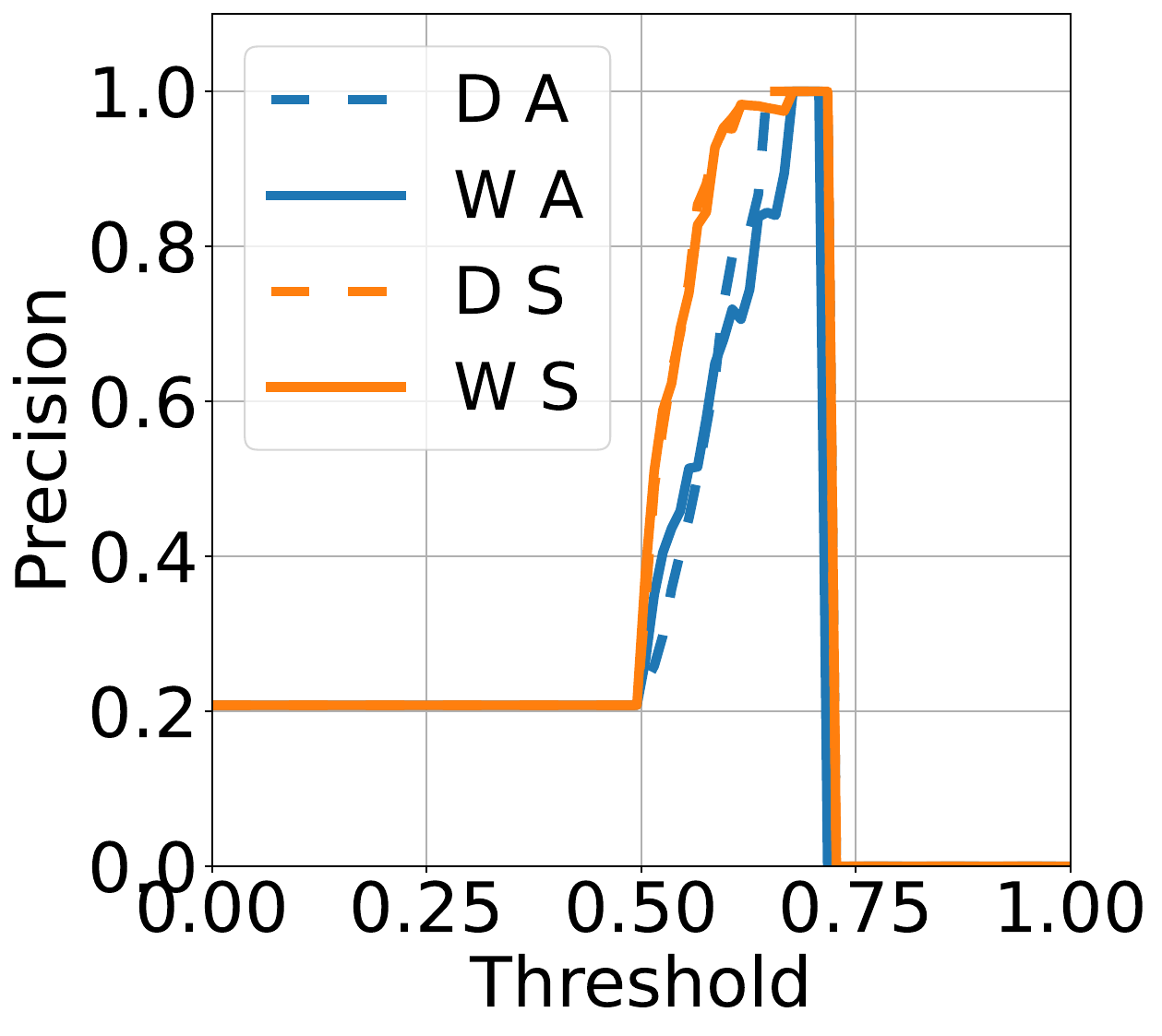}
    \subcaption{Precision.}
    \label{subfig:precision}
  \end{subfigure}\hfill
  \begin{subfigure}[b]{0.19\linewidth}
    \centering
    \includegraphics[width=\linewidth]{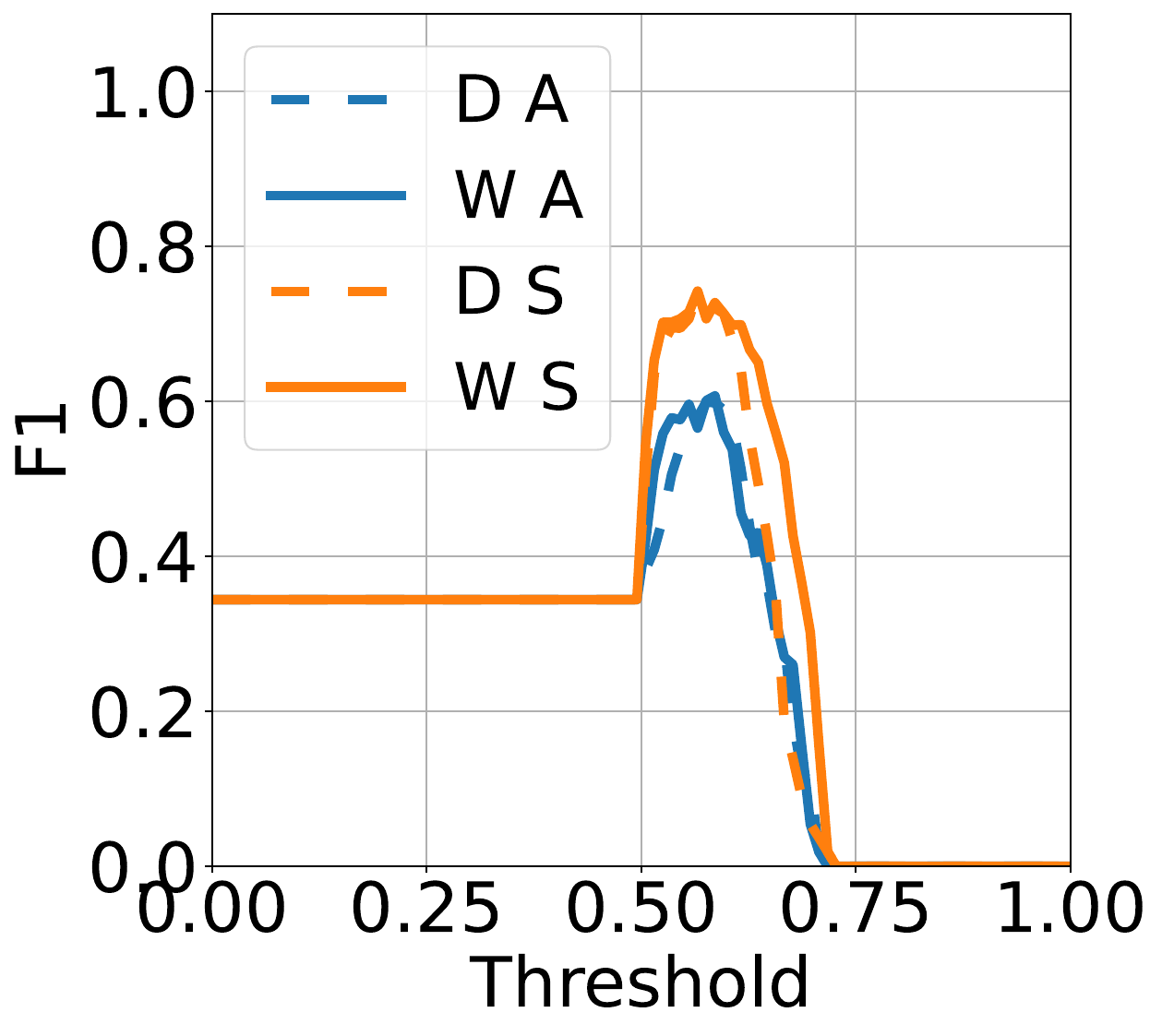}
    \subcaption{F1.}
    \label{subfig:f1}
  \end{subfigure}\hfill
  \begin{subfigure}[b]{0.19\linewidth}
    \centering
    \includegraphics[width=\linewidth]{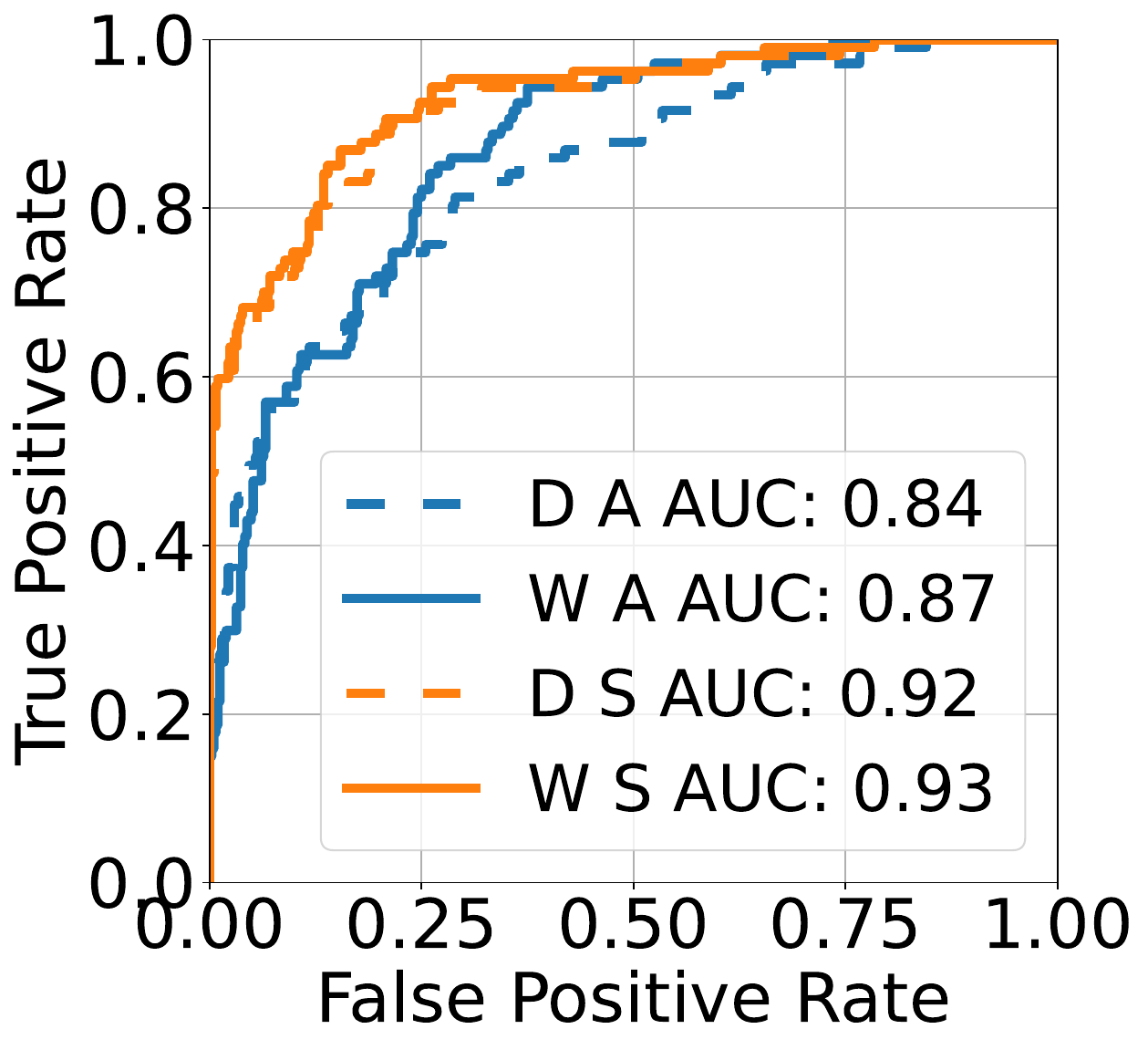}
    \subcaption{AUC.}
    \label{subfig:roc}
  \end{subfigure}\hfill
  \begin{subfigure}[b]{0.19\linewidth}
    \centering
    \includegraphics[width=\linewidth]{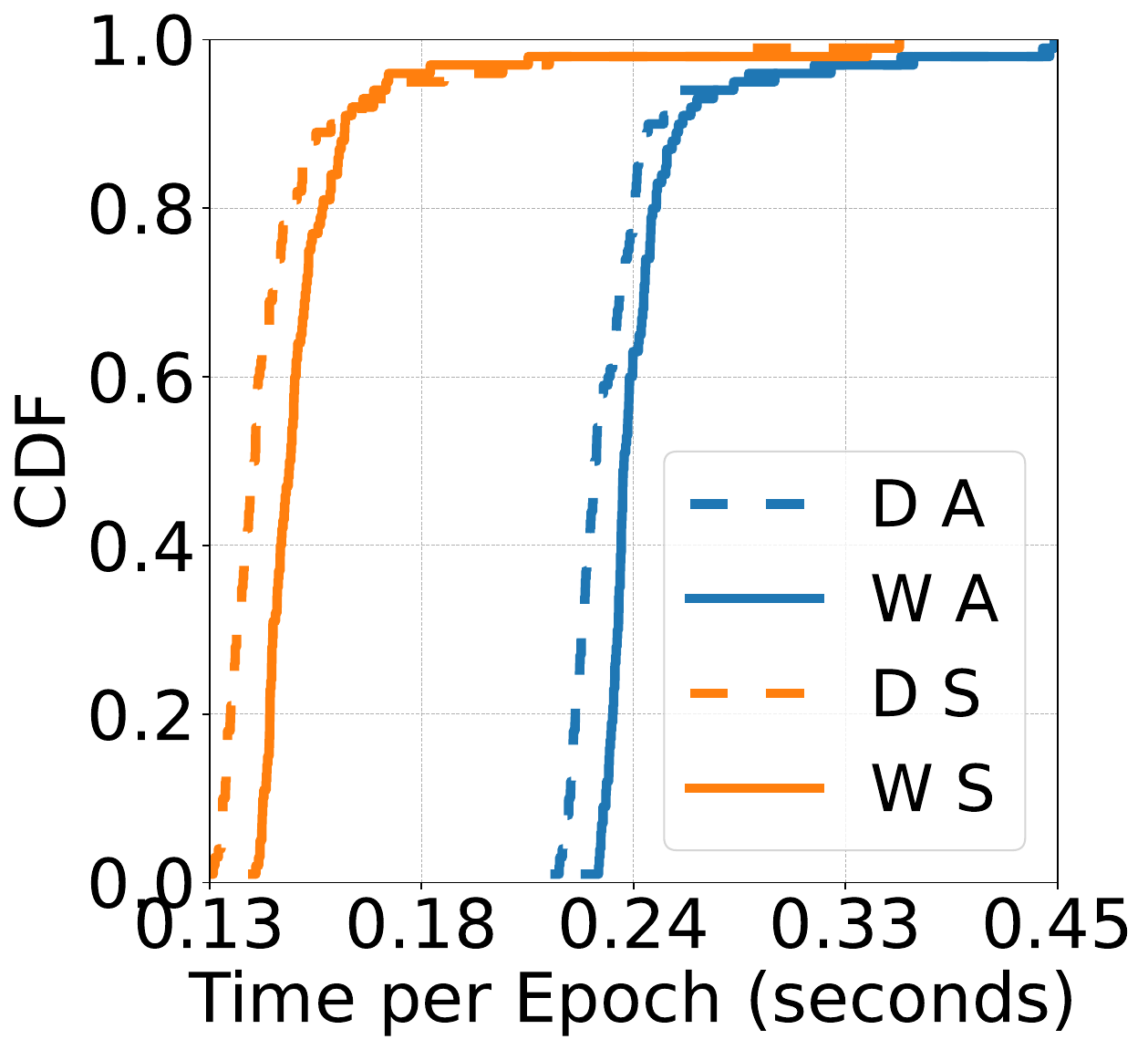}
    \subcaption{Training Speed.}
    \label{subfig:cdf}
  \end{subfigure}\hfill
  \begin{subfigure}[b]{0.19\linewidth}
    \centering
    \includegraphics[width=\linewidth]{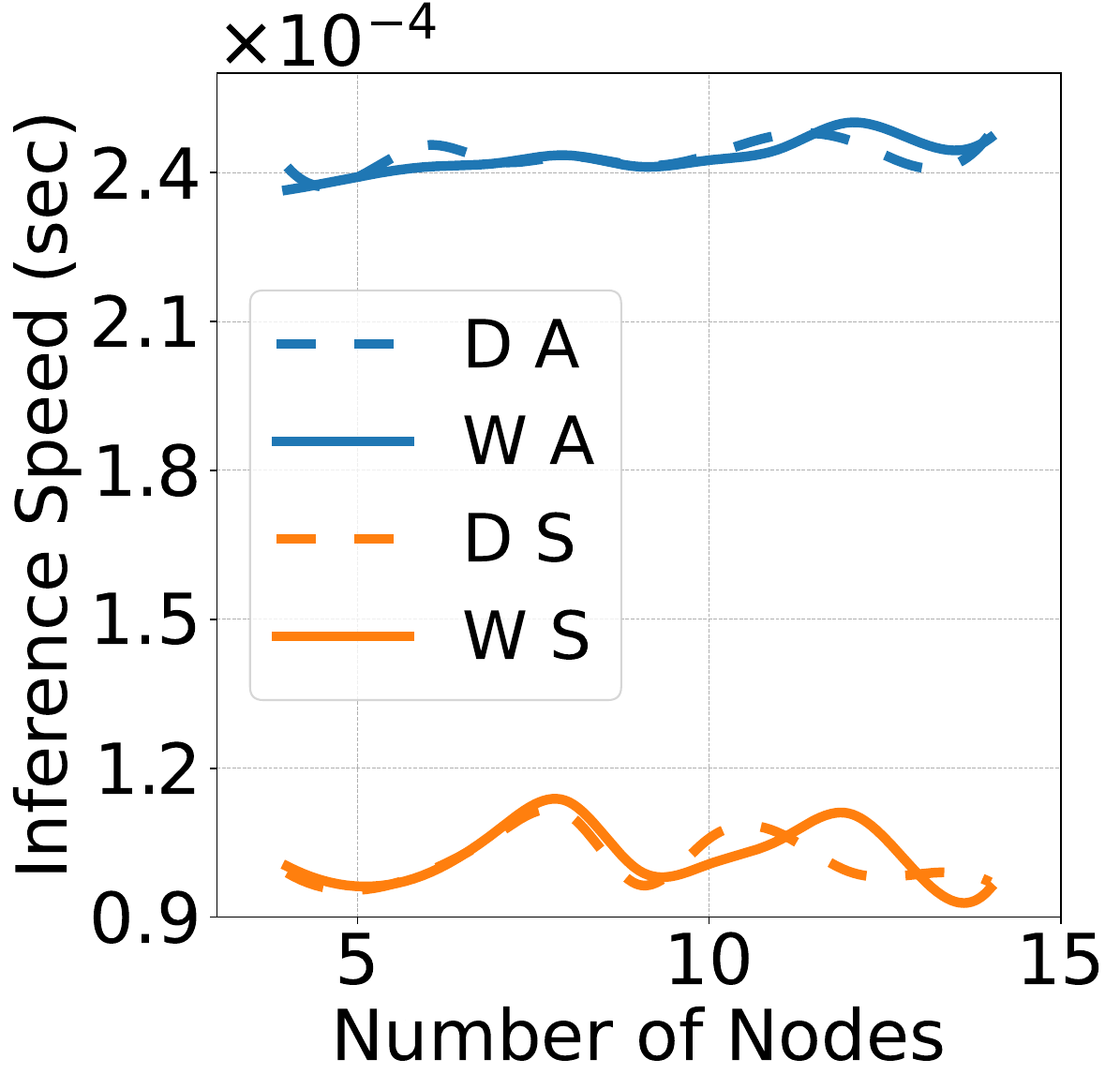}
    \subcaption{Inferring Speed.}
    \label{subfig:inference}
  \end{subfigure}
\end{minipage}
\caption{Efficacy and efficiency plots. D: directed diffusion. W: weighted diffusion. A: GAT. S: GraphSAGE.}
\label{fig:evaluation_five}
\end{figure*}

This section evaluates \textsc{Perseus}' performance, real-world deployment, and detected mastermind characteristics, followed by two case studies illustrating its detection intuitively. Finally, evasion possibility examines mastermind evasion strategies and our countermeasure.

\subsection{Test Scope}
\label{sec: test scope}
To deploy \textsc{Perseus} to detect crowd-pump masterminds in the real world, we collaborate with Cloudburst \cite{cd} to scrape Telegram \ac{osn} messages and monitor real-time cryptocurrency market transactions from Kaiko \cite{kaiko}. We then test \textsc{Perseus} using the real-world data the real-time fetcher collected from April 2018 to October 2024. In total, \textsc{Perseus} supervises $2,103$ channels and monitors $660$ cryptocurrency market transactions. From the $27,365,232$ messages it scrapped, \textsc{Perseus} extracts $733,128$ crowd-pump-related messages through NER. We focus on messages with valid market transactions to ensure relevance, resulting in $49,235$ unique messages. 

For model training and evaluation, we select two timestamps to recursively split real-world data chronologically into training, validation, and testing sets. This process continues until we achieve an approximate distribution of 70\%, 15\%, and 15\% by the number of tokens, with each token being targeted at least by four spreaders. The process splits the data into three periods of time: April 13 2018 to January 9 2024, January 10 to February 4 2024, and February 5 to February 16 2024.

We establish the crowd-pump events for each period, construct temporal attributed graphs, and identify masterminds. \autoref{tab:event_signals} presents a summary statistics of crowd-pump data. Note that the number of graphs does not match the number of cryptocurrencies because some messages are too separated to form crowd-pump events, or the graphs do not have at least four spreaders. In each dataset, the percentage of labeled masterminds varies. Specifically, they are 13.0\% (489) in the training set, 21.4\% (110) in the validation set, and 24.5\% (136) in the test set. The ratio of masterminds in the validation and test set is higher because they cover shorter periods that do not allow enough time for masterminds to spread information to more accomplices.
\begin{table}[b]
\centering
\caption{Summary statistics for real-world crowd-pump data.}
\resizebox{\linewidth}{!}{

\footnotesize
\begin{tabular}{lrrr}
\toprule
& \textbf{04-13-2018 to 01-09-2024} & \textbf{01-10-2024 to 02-04-2024} & \textbf{02-05-2024 to 02-16-2024} \\
& \textbf{Train} & \textbf{Validate} & \textbf{Test} \\
\midrule
Cryptocurrency & 358 & 208 & 208 \\
Crowd-pump & 41,329 & 4,447 & 3,459 \\
Crowd-pump event & 3,523 & 304 & 274 \\
Graph & 306 & 81 & 73 \\
Mastermind & 489 & 110 & 136 \\
Accomplice & 4,300 & 427 & 441 \\
Node & 3,764 & 515 & 554 \\
Directed edge & 7,893 & 1,610 & 1,786 \\
Weighted edge & 8,601 & 1,633 & 1,788 \\
\bottomrule
\end{tabular}
}

\label{tab:event_signals}
\end{table}

\subsection{Evaluation Setup}
\label{sec:evaluation setup}

We evaluate \textsc{Perseus} based on the F1 score, precision, accuracy, recall, and the \ac{roc}. We use \ac{mcc} for cases with close performance due to its previous usage in cryptocurrency financial fraud~\cite{Agarwal2021DetectingProperties, Zola2019CascadingAnonymity}. We evaluate the cumulative density function of the training time and the time to infer each epoch for the detection speed.

We benchmark \textsc{Perseus} with the approach described in \cite{Jiang2019}, which focuses on identifying critical actors in insider threat and financial fraud using \acs{gcn} and a graph construction method integrating direct linkages and cosine similarity between nodes.

We create two adjacency matrices in the Networkx package to represent graph construction methods, directed diffusion and weighted diffusion. We then transform the adjacency matrices into edge lists, combined with node features and one-dimensional labels to form PyTorch geometric objects as inputs for the mastermind detector module. The entire graph for a cryptocurrency forms a batch. We implement \ac{gat} and \ac{sage} with a sigmoid activation function, and then we apply thresholds to obtain binary classifications. 

The experimental setup is an Apple M2 Max CPU with 32GB memory and Python 3.10. We adjust the embedding dimensions and learning rate to optimize the system's performance. For each setting, the experiments are run for 100 epochs. The training and validation phases use the Adam optimizer and employ Binary Cross Entropy as the loss function.

\subsection{Experiment Results}
In this section, we first evaluate different combinations of graph construction methods and \ac{gnn} architectures. After identifying the optimal combination, we fine-tune the model using various parameter settings. Once we determine the best-performing configuration, we benchmark it against the \acs{sota} detection method. Finally, we deploy the model in a the real world and analyze the characteristics of the detected masterminds. To illustrate its algorithm, we present two case studies based on real-world detection results.

\subsubsection{Model Efficacy and Efficiency}
In \autoref{fig:evaluation_five}, to measure efficacy, we plot F1, precision, accuracy, recall, and \acs{roc}, and to measure efficiency. we plot training and inferring speeds.

Precision measures the proportion of correctly predicted positives among all positive predictions. As shown in \autoref{subfig:precision}, \ac{sage} outperforms \ac{gat} in precision within the 0.50–0.6 threshold range, with similar performance elsewhere. The F1 score, which balances precision and recall, follows a similar trend, with \ac{sage} exceeding \ac{gat} between 0.5 and 0.7 (\autoref{subfig:f1}). Both models show a sharp rise in precision (around 0.55) and F1 score (around 0.5), suggesting a natural grouping of spreaders sharing similar characteristics that make them more easily distinguishable by the models at specific thresholds. The \acs{roc} further confirms \ac{sage}'s superior classification ability across false positive rates as shown in \autoref{subfig:roc}. Overall, \ac{sage} proves more effective than \ac{gat} for mastermind detection. Weighted diffusion graphs also outperform directed ones, with weighted diffusion \ac{sage} achieving the highest \acs{roc} (0.93 vs. 0.92 for directed), and weighted diffusion \ac{gat} (0.87) surpassing its directed counterpart (0.84). Thus, weighted diffusion \ac{sage} is the best model.


\autoref{subfig:cdf} indicates the proportion of epochs completed within a specific time threshold. \ac{sage} models take approximately 0.13 to 0.18 seconds to train each epoch, while the \ac{gat} models take more time, around 0.23 to 0.36 seconds. \autoref{subfig:inference} plots the inference speed of models in different node sizes for a batch. Across all node sizes, the \ac{gat} and \ac{sage} maintain a stable inference speed, while the \ac{sage} architectures are faster. Together, \autoref{subfig:cdf} and \autoref{subfig:inference} demonstrate that the \ac{sage} architecture is more efficient in mastermind detection, while the difference between directed and weighted diffusion is negligible. Importantly, it takes less than a second to train and infer a cryptocurrency crowd-pump information diffusion graph, indicating that \textsc{Perseus} can detect masterminds in the fast-changing cryptocurrency market.

\subsubsection{Model Tuning}
\begin{figure}[!t]
  \centering
  \includegraphics[width=\linewidth]{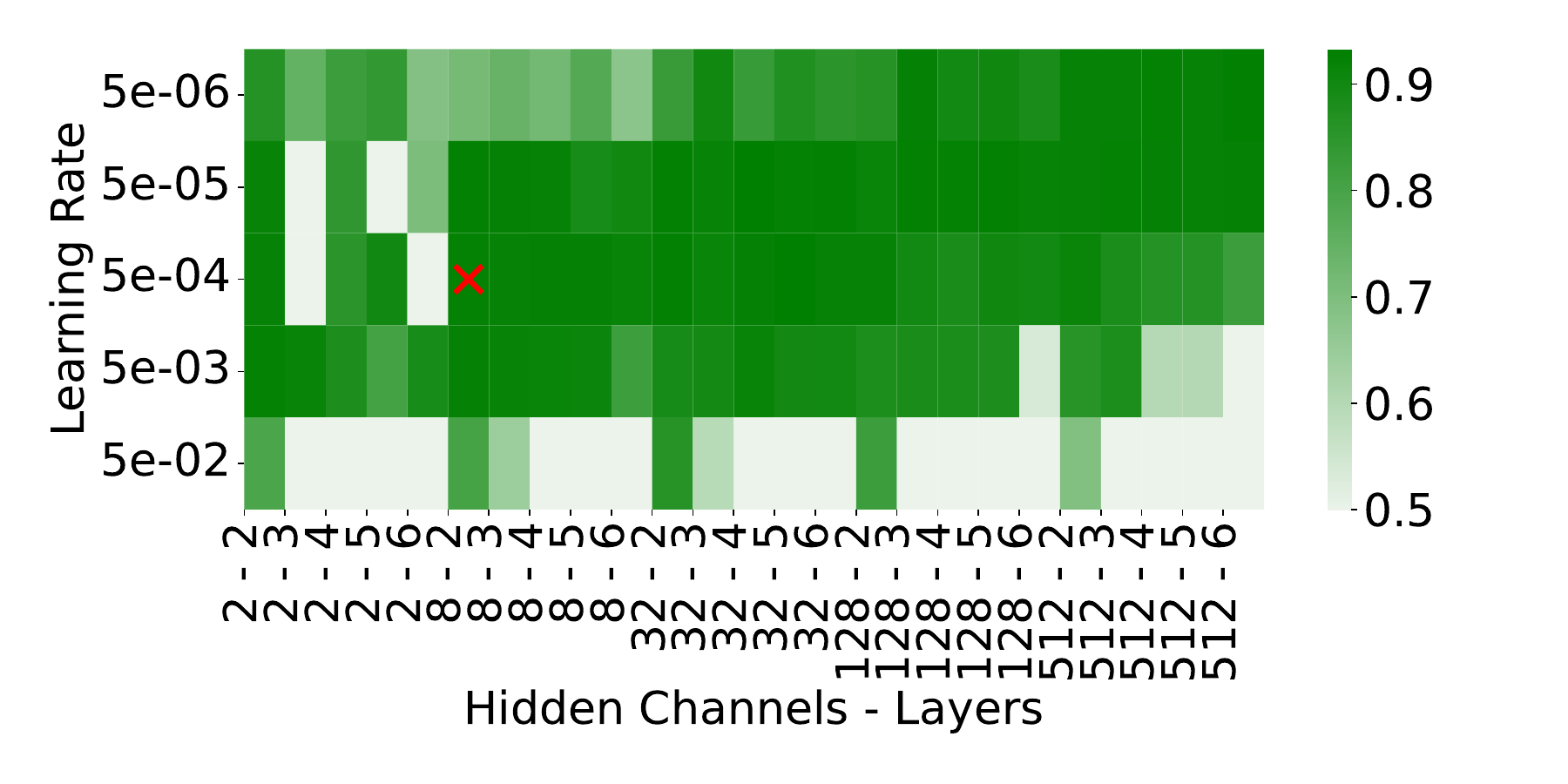}
  \caption{AUC scores for different parameters. In the heat maps, dark green indicates better performance. The cross marks the best parameter setting.}
  \label{fig:auc_scores}
\end{figure}

Our assessments of accuracy, training, and inferring speed show that the weighted diffusion \ac{sage} model is more robust in the mastermind detection task, so we tune its parameters accordingly. We experiment with varying hidden channels at 2, 8, 32, 128, and 512, the number of layers at 2, 3, 4, 5, and 6, and learning rates at 0.05, 0.005, 0.0005, 0.00005, and 0.000005. \autoref{fig:auc_scores} shows that the models at 0.000005 and 0.05 learning rates are not as good as the rest. \autoref{fig:auc_scores} also shows that the models don't perform well for 2 or 512 hidden channels, indicating that either the large or small model in terms of the hidden channels performs well. Lastly, by looking at the number of layers for the model at a learning rate of 0.05, we find that the model performs better at two layers. For best performance, lighter computational load, and avoidance of corner cases, we choose 8 hidden channels, 2 layers, and a learning rate of 0.0005 as our final model. 

\subsubsection{Comparing With the \acs{sota} Detection Method}
 We assess \textsc{Perseus} against the existing method proposed by Jiang~\cite{Jiang2019}. Since crowd-pump spreaders, in our case, do not exhibit directly linked relationships, we adapt Jiang's approach by exclusively using the cosine similarity of node features to construct the graphs. In addition, we also apply random forest as a comparison as it is popular in the crypto pump-and-dump event detection~\cite{Xu2019, LaMorgia2020, Mirtaheri2021}. \autoref{tab:model_performance} shows that weighted diffusion \ac{sage} achieves the highest precision (80.0\%), F1 score (75.2\%), accuracy (90.3\%), and MCC (69.4\%), though with slightly lower recall (71.0\%). The directed diffusion GraphSAGE follows closely, while \ac{gat} models underperform. Weighted diffusion \ac{sage} demonstrates superior mastermind detection, validating \textsc{Perseus}' effectiveness.
\begin{table}[!t]
\centering
\caption{Performance comparison of various models.}
\resizebox{\linewidth}{!}{

\footnotesize
\begin{tabular}{lccccc}
\toprule
Model &  precision &       F1 &  accuracy &   recall & MCC\\
\midrule

       Weighted diffusion GraphSAGE &     \textbf{0.800} & \textbf{0.752} &    \textbf{0.903} &  0.710 & \textbf{0.694} \\
     Directed diffusion GraphSAGE &     0.757 & 0.743 &    0.895 &  0.729 & 0.677 \\       
       Directed diffusion GAT &     0.568 & 0.642 &    0.829 &  \textbf{0.738} & 0.540 \\                    Weighted diffusion GAT &     0.646 & 0.664 &    0.856 &  0.682 & 0.573 \\ \acs{sota} with directed diffusion &     0.768 & 0.667 &    0.878 &  0.589 & 0.601 \\  \acs{sota} with weighted diffusion &     0.694 & 0.698 &    0.874 &  0.701 & 0.618 \\                             Random Forest with weighted diffusion &     0.800 & 0.731 &    0.897 &  0.673 & 0.672 \\
        Random Forest with directed diffusion &     0.738 & 0.724 &    0.887 &  0.710 & 0.653 \\


\bottomrule
\end{tabular}%
}

\label{tab:model_performance}
\end{table}




\begin{table}[!b]
\centering
\caption{T-statistical Test for masterminds and accomplices. *** marks significant at 0.001, ** marks significant at 0.01.}
\resizebox{\linewidth}{!}{

\footnotesize

\begin{tabular}{cc}
\begin{tabular}{lrr}
\toprule
Metric & Statistic & P-value \\
\midrule
Clustering coefficient & 37.5\textsuperscript{***} & 3.12e-176 \\
Betweenness centrality & -22.0\textsuperscript{***} & 1.37e-73 \\
Closeness centrality & 3.79\textsuperscript{***} & 5.65e-4 \\
Effective size & -30.2\textsuperscript{***} & 1.31e-109 \\
Out degree & -28.6\textsuperscript{***} & 1.71e-103 \\
\bottomrule
\end{tabular}
&
\begin{tabular}{lrr}
\toprule
Metric & Statistic & P-value \\
\midrule
Pagerank & -3.96\textsuperscript{***} & 1.36e-4 \\
Out ratio & -47.8\textsuperscript{***} & 6.41e-188 \\
In ratio & 2.9\textsuperscript{**} & 5.50e-3 \\
Efficiency & -28.0\textsuperscript{***} & 1.17e-147 \\
Density & 16.5\textsuperscript{***} & 1.69e-59 \\
\bottomrule
\end{tabular}
\end{tabular}
}

\label{tab:mastermind}
\end{table}

\subsubsection{Real-world detection}
We deploy \textsc{Perseus} in real-world and report its performance from February 16 to October 9 2024. During that period, masterminds significantly impact the market, driving a $167\%$ increase in trading volumes. We define crowd-pump duration as the period within three days from the time of announcement to when the price peaks, and the total crowd-pump trading volumes as the sum of volumes traded during this period. Regular trading volumes are estimated by calculating the average trading volume per minute over the three days before the crowd-pump announcement and multiplying it by the crowd-pump duration. The total crowd-pump trading volumes reached \$8.07 trillion, compared to an estimated \$4.83 trillion in regular trading, increasing by \$3.24 trillion ($67\%$).

During the period, \textsc{Perseus} identifies 438 masterminds associated with 322 distinct cryptocurrencies. The distribution of masterminds across cryptocurrencies varies. In particular, 192 cryptocurrencies are targeted by only one mastermind, 72 are targeted by two masterminds, and 23 are linked to three masterminds. Remarkably, four masterminds target the following cryptocurrencies: \textit{AAVE}, \textit{ATOM}, \textit{BAKE}, \textit{BCH}, \textit{COTI}, \textit{FIL}, and \textit{LINK}. The most frequently targeted cryptocurrency is \textit{BTC} with five masterminds identified.

\begin{figure}[!t] 
    \centering
    \begin{minipage}{0.49\linewidth}
        \includegraphics[width=\linewidth]{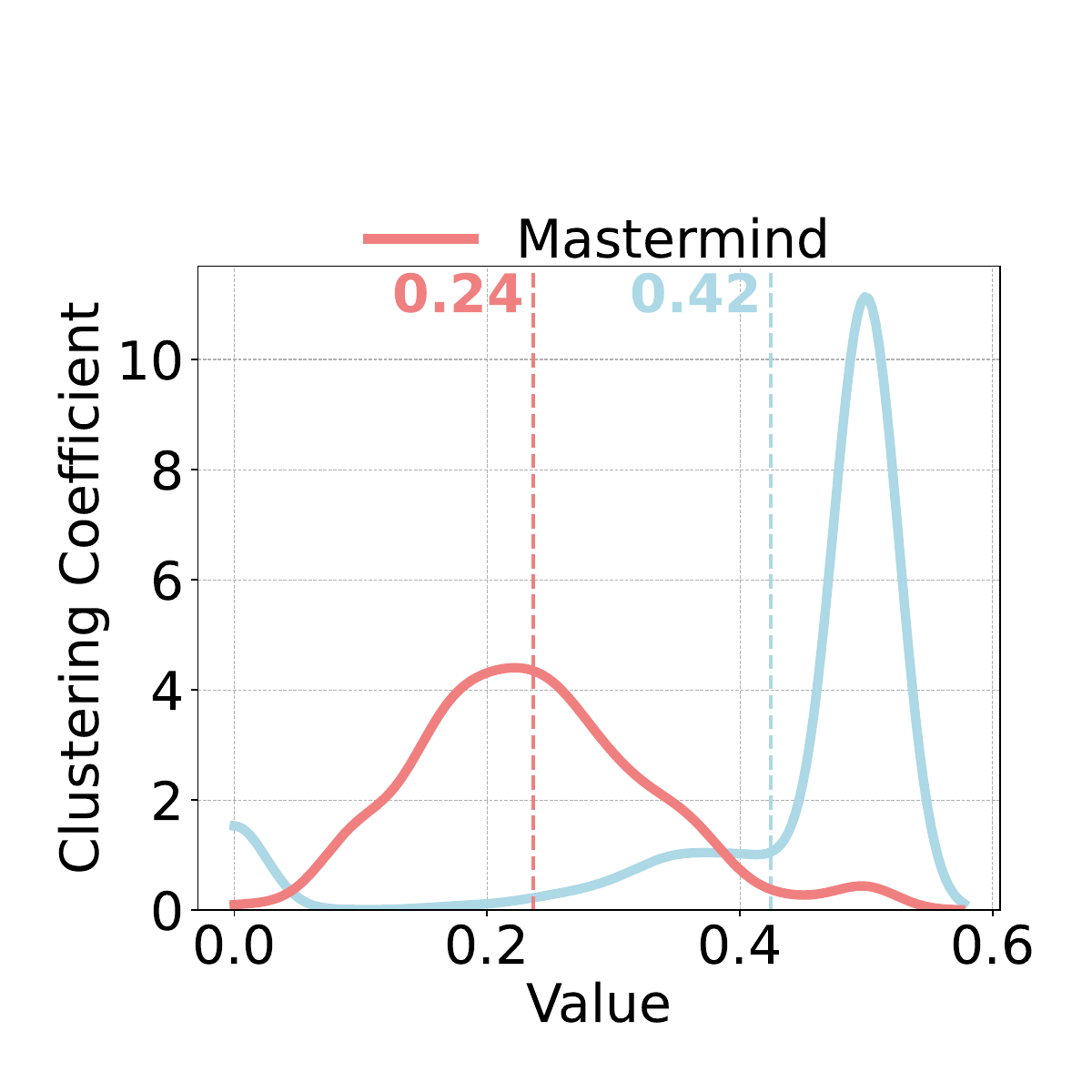}
        \subcaption{Probability Density Distribution for Clustering Coefficient.}
        \label{fig:Clustering_Coefficient}
    \end{minipage}%
    \hfill
    \begin{minipage}{0.49\linewidth}
        \includegraphics[width=\linewidth]{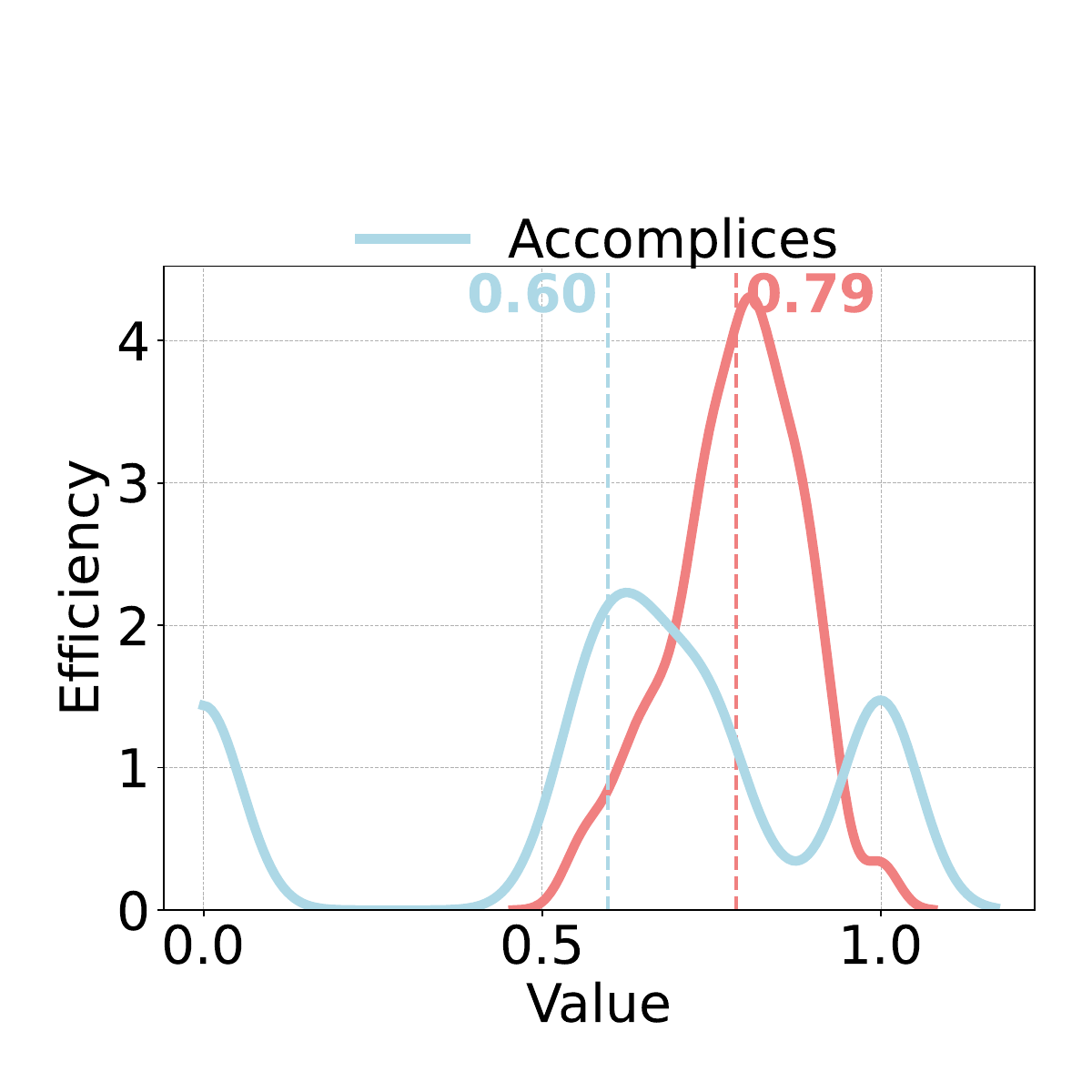}
        \subcaption{Probability Density Distribution for Efficiency.}
        \label{fig:Efficiency}
    \end{minipage}
    \caption{Distributions of topological features. The dashed vertical lines indicate the mean.}
    \label{fig:distribution}
\end{figure}

Furthermore, we conducted a t-test on the topological features for masterminds and accomplices on the directed diffusion graphs and documented the results in \autoref{tab:mastermind}. The t-test is used to determine whether there is a significant difference between the means of the two groups. \autoref{tab:mastermind} shows distinct network characteristics between masterminds and accomplices. We select the two significant features and plot their probability density distributions in \autoref{fig:distribution}. \autoref{fig:Clustering_Coefficient} shows that the accomplices have a higher mean clustering coefficient at 0.42 with a variance of 0.02, indicating that they operate within tightly connected local clusters and propagate information within. In \autoref{fig:Efficiency}, mastermind demonstrates a sharply concentrated distribution around a high mean efficiency of 0.84 with a minimal variance of 0.02. This shows that the mastermind is superiorly efficient in their ego information diffusion network, indicating that the accomplices are more likely to get the crowd-pump information directly from masterminds than their peer accomplices.

\subsubsection{Case Study}
\begin{figure}[!t] 
    \centering
    \begin{subfigure}{0.49\linewidth}
        \includegraphics[width=\linewidth]{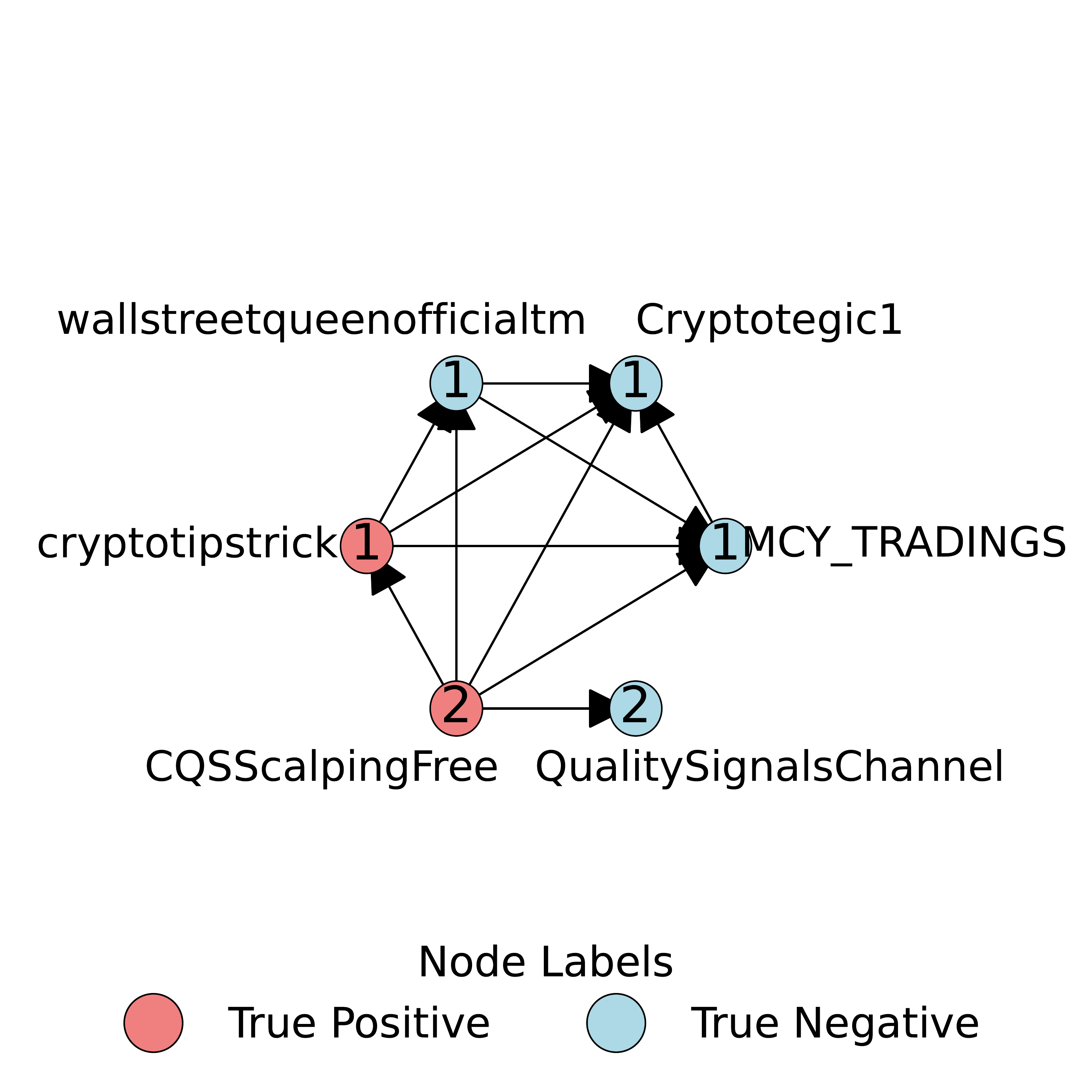}
        \subcaption{Case \textit{SUI}.}
        \label{fig:SUI}
    \end{subfigure}%
    \hfill
    \begin{subfigure}[b]{0.49\linewidth}
        \includegraphics[width=\linewidth]{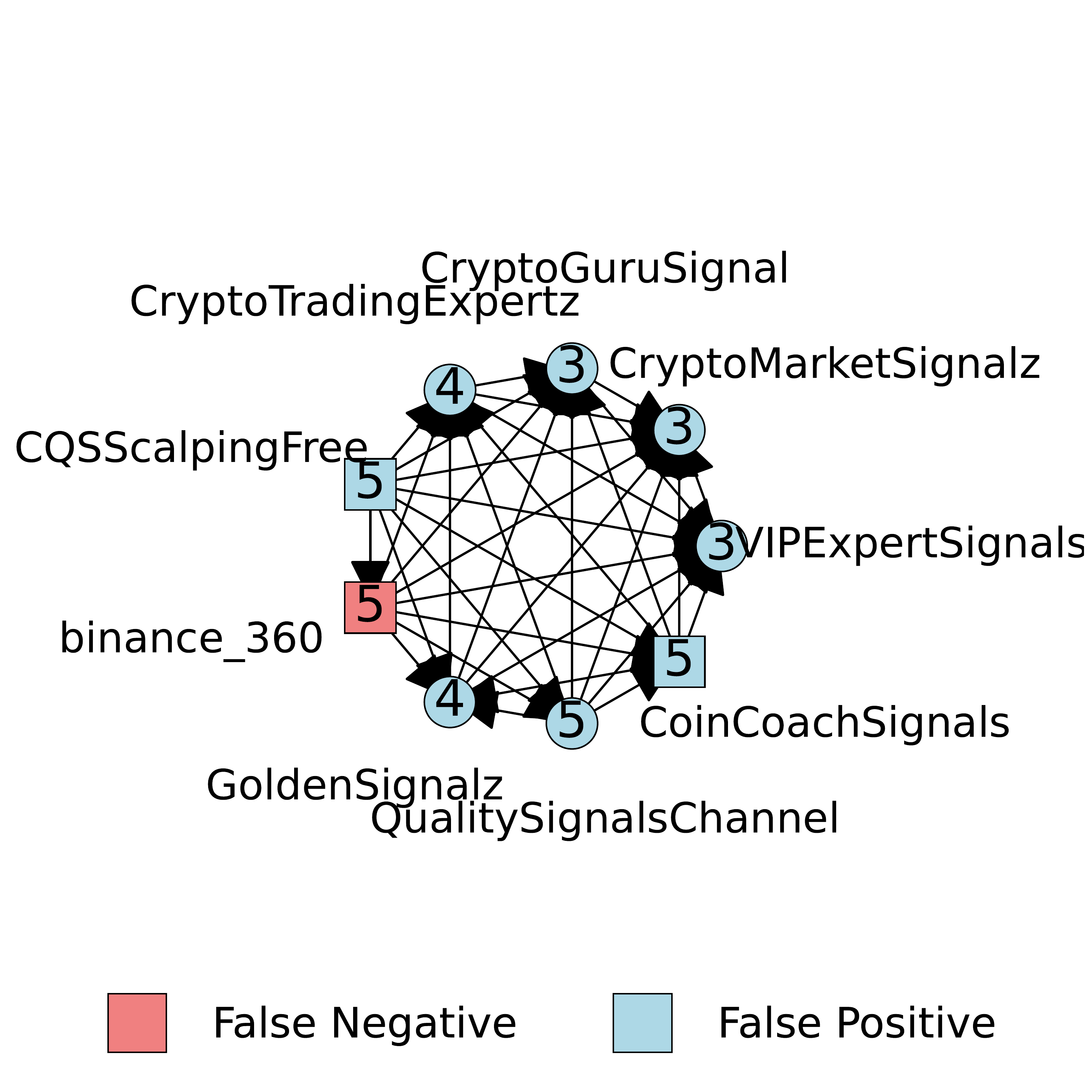}
        \subcaption{Case \textit{STORJ}.}
        \label{fig:STORJ}
    \end{subfigure}
    \caption{Case studies illustration. The edges come from directed diffusion graph. Community are inferred from weighted diffusion graph. Numbers represent different communities. }
    \label{fig:cases}
\end{figure}

In \autoref{fig:cases}, we apply the Louvain community detection algorithm on the weighted diffusion graph to infer communities and plot the main direction of information diffusion using the directed diffusion graph. 
\paragraph{Case \textit{SUI}}
\autoref{fig:case_sui_crop} illustrates how crowd-pumps on \textit{SUI} impact the market. The detailed crowd-pump messages can be found in \autoref{tab:SUI}. We find two masterminds in the crowd-pump events on \textit{SUI} from February 6 to February 13 2024. \textit{SUI} is the native token built for a layer one blockchain optimizing for low latency~\cite{sui}. There are two masterminds in \textit{SUI} crowd-pumps, cryptotipstrick and CQSScalpingFree. From \autoref{fig:SUI}, in community 2, CQSScalpingFree's edges are all outward, meaning that the mastermind only distributes the information rather than receives it. This allows \textsc{Perseus} to identify it as the mastermind. Connected to two communities, cryptotipstrick has three outward edges in community 1, allowing \textsc{Perseus} to detect it as the mastermind.

\paragraph{Case \textit{STORJ}}

From the investigation in \autoref{case_study_2}, we find two masterminds in the crowd-pump events on \textit{STORJ} from February 13 to February 14 2024. \textit{STORJ} is a cryptocurrency token used to power the \textit{STORJ} decentralized cloud storage network \cite{stojr}. In \autoref{fig:STORJ}, \textsc{Persues} fails to detect masterminds because the spreaders broadcast the messages simultaneously, making \textsc{Perseus} unable to construct representative graphs of the information diffusion network.

\subsection{Evasion Possibility}
To avoid the detection system, masterminds can utilize the following strategy. 
\begin{enumerate*}[label={(\roman*)}]
    \item \label{strat:i} Node addition: masterminds create off-scope channels, adding a new node \(v_{\text{new}}\) to \(V\), altering the graph \(G\)'s topology;
    \item \label{strat:ii} Edge manipulation: masterminds and accomplices broadcast crowd-pump messages simultaneously, adding misleading edges \(E_{\text{concurrent}}\) to \(E\), distorting the network structure; and
    \item \label{strat:iii} Identity manipulation: masterminds use bots to spread messages, introducing disguised nodes \(V_{\text{bot}}\) into \(V\), concealing their identity without exposing metadata.
\end{enumerate*}


To counteract strategy \ref{strat:i}, we can expand the scope of monitoring by continuously identifying new channels. If an unusual number of crowd-pump messages appear simultaneously or bots are detected as described in strategies \ref{strat:ii} and \ref{strat:iii}, the system can flag them for further human inspection.

\begin{table*}[!h]
\centering
\caption{The crowd-pump event tables for \textit{SUI} and \textit{STORJ}, outlining the channels, timing, specific messages sent, and their corresponding returns. $d_1$, $d_2$, and $d_3$ denote individual crowd-pump events. Red texts highlight the crowd-pump events in which masterminds participate. Blue texts highlight the crowd-pump events in which accomplices participate. Orange texts represent false positive cases. Return is the maximum return achieved within three days after the announcement of the message.}
\footnotesize

\begin{subtable}[t]{\textwidth}
\centering

\footnotesize

\begin{tabularx}{\textwidth}{|c|l|c|X|c|}
\hline
Events & Telegram Channels & Timestamps & Messages & Returns \\
\hline
\textcolor{red}{$d1$} & \textcolor{red}{CQSScalpingFree} & \textcolor{red}{2024-02-06 18:25:40} & \textcolor{red}{BINANCE USDT\_SUI LONG 3415609 Ask 1.53360000 Target 1.55481690 TP 1.4 SL 4.0} &  \textcolor{red}{1.28\%} \\
\textcolor{blue}{} & \textcolor{blue}{QualitySignalsChannel} & \textcolor{blue}{2024-02-08 10:44:14} & \textcolor{blue}{New FREE signal BUY SUIUSDT at BINANCE Leverage SPOT 1x 2491497 08Feb2024 104410 UTC Entry Zone 1.44130249 1.545329 Current ask 1.5346 Target 1 1.60319 4.47 Target 2 1.62418999 5.84 Target 3 1.64518999 7.21 Stop loss 1.40876999 8.20 Volume SUI 48952889.700000 Volume USDT 75163544.712950 SHORT TERM up to 7 days Risk 35 Medium Invest up to 5 of your portfolio RR ratio 0.7 Chart SUIUSDT can also be traded on BINANCE, BINANCEUS, BINANCE\_FUTURES, BITGET, BYBIT, HUOBI, KUCOIN, OKEX, POLONIEX The information here is not financial advice. It comes without any guarantees. Invest wisely and manage your risks.} &  \textcolor{blue}{7.25\%} \\
\hline
\textcolor{red}{$d2$} & \textcolor{red}{CQSScalpingFree} & \textcolor{red}{2024-02-11 01:05:53} & \textcolor{red}{BINANCE BTC\_SUI LONG 3418235 Ask 0.00003647 Target 0.00003687 TP 1.1 SL 4.0} &  \textcolor{red}{1.04\%} \\
\textcolor{blue}{} & \textcolor{red}{cryptotipstrick} & \textcolor{red}{2024-02-13 18:35:58} & \textcolor{red}{SUIUSDT Exchanges Binance Futures Signal Type Regular Long Leverage Cross 50x Entry Targets 1.8225 TakeProfit Targets 11.85 21.87 31.89 41.91 51.93 61.96 7 Stop Targets 510} &  \textcolor{red}{7.52\%} \\
\textcolor{blue}{} & \textcolor{blue}{wallstreetqueenofficialtm} & \textcolor{blue}{2024-02-13 20:09:16} & \textcolor{blue}{Coin SUIUSDT Exchange Binance Futures Signal Type Regular Long Leverage Cross 50x ENTRY Targets 1.8225 TAKE PROFIT Targets 11.85 21.87 31.89 41.91 51.93 61.96 STOP LOSS Target 510 Binance Pumps} &  \textcolor{blue}{5.95\%} \\
\textcolor{blue}{} & \textcolor{blue}{MCY\_TRADINGS} & \textcolor{blue}{2024-02-13 20:13:17} & \textcolor{blue}{SUIUSDT Exchanges Binance Futures Signal Type Regular Long Leverage Cross 50x Entry Targets 1.8225 TakeProfit Targets 11.85 21.87 31.89 41.91 51.93 61.96 7 Stop Targets 510} &  \textcolor{blue}{6.06\%} \\
\textcolor{blue}{} & \textcolor{blue}{Cryptotegic1} & \textcolor{blue}{2024-02-13 20:14:28} & \textcolor{blue}{SUIUSDT Exchanges Binance Futures Signal Type Regular Long Leverage Cross 50x Entry Targets 1.8225 TakeProfit Targets 11.85 21.87 31.89 41.91 51.93 61.96 7 Stop Targets 510} &  \textcolor{blue}{5.95\%} \\
\hline
\end{tabularx}
\caption{Case \textit{SUI}.}
\label{tab:SUI}
\end{subtable}

\vspace{10pt}

\begin{subtable}[t]{\textwidth}
\centering

\footnotesize

\label{tab: case_study_wrong}
\begin{tabularx}{\textwidth}{|c|l|c|X|c|}
\hline
Events & Telegram Channels & Timestamps & Messages & Returns \\
\hline
\textcolor{red}{$d3$} & \textcolor{red}{CQSScalpingFree} & \textcolor{red}{2024-02-13 10:44:46} & \textcolor{red}{BINANCE BTC\_STORJ LONG 3419859 Ask 0.00001347 Target 0.00001362 TP 1.1 SL 4.0} &  \textcolor{red}{1.11\%} \\
\textcolor{orange}{} & \textcolor{orange}{binance\_360} & \textcolor{orange}{2024-02-14 08:00:24} & \textcolor{orange}{Futures Free Signal Long STORJUSDT Entry zone 0.6466\_0.6598 Targets 0.6643\_0.6775\_0.6907\_0.7038\_0.7170\_0.7301\_0.7433\_0.7564 Stop loss 0.6136 Leverage 5x\_10x binance\_360} &  \textcolor{orange}{10.72\%} \\
\textcolor{blue}{} & \textcolor{blue}{QualitySignalsChannel} & \textcolor{blue}{2024-02-14 12:58:30} & \textcolor{blue}{New FREE signal BUY STORJUSDT at BINANCE\_US Leverage SPOT 1x 2492909 14Feb2024 125304 UTC Entry Zone 0.62734751 0.664945 Current ask 0.6606 Target 1 0.67885 2.76 Target 2 0.68674001 3.96 Target 3 0.69463001 5.15 Stop loss 0.61533001 6.85 Volume STORJ 20909.000000 Volume USDT 13786.841400 SHORT TERM up to 7 days Risk 35 Medium Invest up to 5 of your portfolio RR ratio 0.6 Chart STORJUSDT can also be traded on BINANCE, BINANCEUS, BINANCE\_FUTURES, BITGET, HUOBI, KUCOIN, OKEX, POLONIEX The information here is not financial advice. It comes without any guarantees. Invest wisely and manage your risks.} &  \textcolor{blue}{5.29\%} \\
\textcolor{red}{} & \textcolor{red}{CoinCoachSignals} & \textcolor{red}{2024-02-14 13:07:17} & \textcolor{red}{CAPTION STORJUSDT LONG ENTRY 0.6620 0.6508 Leverage Cross 10X TAKE PROFIT 1 0.6750 2 0.6827 3 0.6950 4 0.7059 5 0.7155 Stop Loss 0.6235} &  \textcolor{red}{8.05\%} \\
\textcolor{blue}{} & \textcolor{blue}{GoldenSignalz} & \textcolor{blue}{2024-02-14 13:07:17} & \textcolor{blue}{CAPTION STORJUSDT LONG ENTRY 0.6620 0.6508 Leverage Cross 10X TAKE PROFIT 1 0.6750 2 0.6827 3 0.6950 4 0.7059 5 0.7155 Stop Loss 0.6235} &  \textcolor{blue}{8.05\%} \\
\textcolor{blue}{} & \textcolor{blue}{CryptoTradingExpertz} & \textcolor{blue}{2024-02-14 13:07:17} & \textcolor{blue}{CAPTION STORJUSDT LONG ENTRY 0.6620 0.6508 Leverage Cross 10X TAKE PROFIT 1 0.6750 2 0.6827 3 0.6950 4 0.7059 5 0.7155 Stop Loss 0.6235} &  \textcolor{blue}{8.05\%} \\
\textcolor{blue}{} & \textcolor{blue}{CryptoGuruSignal} & \textcolor{blue}{2024-02-14 13:07:17} & \textcolor{blue}{CAPTION STORJUSDT LONG ENTRY 0.6620 0.6508 Leverage Cross 10X TAKE PROFIT 1 0.6750 2 0.6827 3 0.6950 4 0.7059 5 0.7155 Stop Loss 0.6235} &  \textcolor{blue}{8.05\%} \\
\textcolor{blue}{} & \textcolor{blue}{VIPExpertSignals} & \textcolor{blue}{2024-02-14 13:07:17} & \textcolor{blue}{CAPTION STORJUSDT LONG ENTRY 0.6620 0.6508 Leverage Cross 10X TAKE PROFIT 1 0.6750 2 0.6827 3 0.6950 4 0.7059 5 0.7155 Stop Loss 0.6235} &  \textcolor{blue}{8.05\%} \\
\textcolor{blue}{} & \textcolor{blue}{CryptoMarketSignalz} & \textcolor{blue}{2024-02-14 13:07:17} & \textcolor{blue}{CAPTION STORJUSDT LONG ENTRY 0.6620 0.6508 Leverage Cross 10X TAKE PROFIT 1 0.6750 2 0.6827 3 0.6950 4 0.7059 5 0.7155 Stop Loss 0.6235} &  \textcolor{blue}{8.05\%} \\
\hline
\end{tabularx}
\caption{Case \textit{STORJ}.}
\label{case_study_2}
\end{subtable}

\end{table*}

\section{Related work}
\label{sec:relatedwork}


\begin{table}[!t]
\centering
\caption{Related work comparison table shows that our study is the first work on mastermind detection with topological features.}
\footnotesize
\resizebox{\linewidth}{!}{%

\begin{tabular}{lcccccc}
      \toprule  
      \multirow{2}{*}{Reference}                            & \multirow{2}{*}{Methods}      & \multicolumn{3}{c}{Features}       & \multirow{2}{*}{Crowd-pump}        & \multirow{2}{*}{Mastermind Detection}  \\
      \cmidrule(lr){3-5}

                                           &       & Market               & \ac{osn}            & Topology            &                              &                       \\
      \midrule
      \cite{Xu2019}                        & Random Forest                                                                          & \CIRCLE                 & \Circle                & \Circle                          & \Circle                     & \Circle              \\
      \cite{LaMorgia2020}                  & Random Forest                                                                                                    & \CIRCLE                 & \CIRCLE                & \Circle                          & \Circle                     & \Circle              \\
      \cite{Chadalapaka2022}               & CLSTM and Anomaly                                                                                 & \CIRCLE                 & \Circle                & \Circle                          & \Circle                     & \Circle              \\
      \cite{Nilsen2019}                    & LTSM                                                     & \CIRCLE                 & \Circle                & \Circle                          & \Circle                     & \Circle              \\
      \cite{Hu2022}                        & SNN                                                                    & \CIRCLE                 & \CIRCLE                & \Circle                          & \Circle                     & \Circle              \\
      \cite{Kamps2018a}                    & Threshold algorithm                                                     & \CIRCLE                 & \Circle                & \Circle                          & \Circle                     & \Circle              \\
      \cite{Chen2019c}                     &  Apriori algorithm                                                                    & \CIRCLE                 & \Circle                & \Circle                          & \Circle                     & \Circle              \\
      \cite{Hamrick2021}                   & Threshold algorithm                                                                      & \CIRCLE                 & \CIRCLE                & \Circle                          & \CIRCLE                     & \Circle              \\
      \cite{Nghiem2021}                    & CNN, BLSTM, and CLSTM                                       & \CIRCLE                 & \CIRCLE                & \Circle                          & \Circle                     & \Circle              \\
      \cite{Victor2019a}                   & XGBoost                                        & \CIRCLE                 & \CIRCLE                & \Circle                          & \Circle                     & \Circle              \\
      \cite{Mirtaheri2021}                 & Random Forest                                     & \CIRCLE                 & \CIRCLE                & \Circle                          & \CIRCLE                     & \Circle              \\
      \textsc{Persues}                              & GNN                                                                     & \CIRCLE                 & \CIRCLE                & \CIRCLE                          & \CIRCLE                     & \CIRCLE              \\
      \bottomrule
\end{tabular}
}

\label{tab:litreview}

\end{table}

\textsc{Persues} is the first study to address pump-and-dump schemes at the root by pinpointing masterminds who are at the source of crowd-pumps, as shown in \autoref{fig:case_sui_crop}. Since 2018 when Kamps et al.~\cite{Kamps2018a} first introduced the cryptocurrency pump-and-dump detection, various studies have used time series analysis, regression, and machine learning~\cite{Nilsen2019, LaMorgia2020, Chadalapaka2022, Xu2019, Hu2022, Victor2019a, Mirtaheri2021} to improve detection, offering solutions at the exchange market level. Hamrick and Nghiem~\cite{Hamrick2021, Nghiem2021} used \ac{osn} signals and market data to predict pump-and-dump success and target cryptocurrencies, highlighting \ac{osn} collaboration but still focusing on the exchange market level. Chen et al.~\cite{Chen2019c} proposed an improved apriori algorithm to detect the major investors, while Yahya~\cite{Yahya2024, Merkley2024Crypto-influencers} examined the spreaders on \ac{osn} without identifying the masterminds. In \autoref{tab:litreview}, we present the comparison of our research with others.

\section{Discussion and Conclusion}
\label{sec:Discussion}

In this research, we passively collect data without interacting with pump-and-dump schemes. Once masterminds are identified, their publicly available information (e.g., nicknames, emails, phone numbers) can be shared with authorities. Since we focus on identifying major entities across \ac{osn}s, our study raises no ethical disclosure concerns.

However, limitations exist. Some pump events may lack labeled masterminds due to undemonstrative influence, and discrepancies between reality and human labeling remain, as true masterminds are often unknown. Additionally, detection is time- and token-specific, and we extract only key information from each message. Future work could aggregate pump events across tokens and leverage large language models which have shown strong results in various domains.


\section{Conclusion}
\label{sec:Conclusion}

Pump-and-dump schemes in the cryptocurrency market pose a significant threat to investors. Our research is the first to pinpoint the masterminds behind such schemes and document their characteristics. First, we gather data from the \acs{osn}s and cryptocurrency markets. Then we construct temporal attributed graphs to capture the direction of information diffusion and community structure among spreaders. Subsequently, we develop a \acs{gnn} network to detect masterminds and provide explanation on their risks. Our research detects mastermind efficiently and effectively and finds them responsible for passing the crowd-pump messages directly to accomplices. Leveraging our innovative approaches, regulators can significantly enhance their oversight capabilities, reinforcing the integrity and stability of cryptocurrency markets.

\bibliographystyle{IEEEtran}
\bibliography{references_pump_dump}

\printacronyms

\end{document}